\documentclass[11pt,aps,showpacs,notitlepage,nofootinbib]{revtex4-1}

\usepackage{amssymb,amsmath}
\usepackage{bm}
\usepackage[pdftex]{graphicx}
\usepackage{subfig}
\usepackage{color}
\usepackage{amsthm}
\usepackage[update,prepend]{epstopdf}
\usepackage{hyperref}
\usepackage{float}

\usepackage{footnote}

\begin{document}

\title{Gravitational waves from spectator Gauge-flation
}
\author{
Oksana~Iarygina$^{1}$, Evangelos~I.~Sfakianakis$^{1,2,3}$
}
\email{\baselineskip 11pt Email addresses:  iarygina@lorentz.leidenuniv.nl; esfakianakis@ifae.es }
\affiliation{
$^1$Institute Lorentz of Theoretical Physics, Leiden University, 2333 CA Leiden, The Netherlands
\\
$^2$ Nikhef, Science Park 105, 1098 XG Amsterdam, The Netherlands
\\
$^3$ Institut de Fisica d’Altes Energies (IFAE), The Barcelona Institute of Science and Technology (BIST), Campus UAB, 08193 Bellaterra, Barcelona
}
\date{\today}

\begin{abstract}
We investigate the viability of inflation with a spectator sector comprised of non-Abelian gauge fields coupled through a higher order operator. We dub this model ``spectator Gauge-flation''.  
We study the predictions for the amplitude and tensor tilt of chiral gravitational waves and
 conclude that 
 a slightly red-tilted tensor power spectrum is preferred with $n_T=-{\cal O}(0.01)$. As with related models, the
 enhancement of chiral gravitational waves with respect to the single-field vacuum gravitational wave background
 is controlled by the parameter $\gamma=g^2 Q^2/H^2$, where $g$ is the gauge coupling, $H$ is the Hubble scale and $Q$ is the VEV of the $SU(2)$ sector.
The requirement that the $SU(2)$ is a spectator sector leads to a maximum allowed value for $\gamma$, thereby constraining the possible amplification.
In order to provide concrete predictions, we use an $\alpha$-attractor T-model potential for the inflaton sector.
  Potential observation of chiral gravitational waves with significantly tilted tensor spectra would then indicate the presence of additional couplings of the gauge fields to axions, like in the spectator axion-SU(2) model, or additional gauge field operators.
\end{abstract}
%\pacs{Preprint numbers}
\maketitle

\newpage
\tableofcontents

\section{Introduction}

Inflation provides an elegant solution for the horizon and flatness problems, as well as a mechanism for producing density fluctuations in very good agreement with the latest observational tests \cite{Aghanim:2018eyx}. Typically, scalar fields play a major role in inflationary model-building since they do not spoil the homogeneity and isotropy of the background cosmology. However, models of particle physics generically include  gauge fields and their presence in the inflationary epoch may significantly influence  cosmological predictions. Scalar perturbations that are produced during inflation are tightly constrained by observations \cite{Aghanim:2018eyx}, while the primordial tensor modes (generated as primordial gravitational waves) are still not detected. The primordial Stochastic Gravitational
Wave Background (SGWB) is a unique test of the physics of the very early Universe, that could provide signatures of the particle content and the energy scale of  inflation. Nowadays, the search for primordial gravitational waves (GWs)  is mainly focused \cite{Abazajian:2020dmr, Campeti:2020xwn} on the parity-odd polarization pattern in the CMB the B-modes. A correct interpretation of B-mode measurements strongly relies on  understanding  their production mechanism.

One intriguing scenario is GW generation by gauge fields. Gauge field tensor modes can experience a tachyonic growth in one of their polarizations,  leading to the production of chiral GWs. In addition to chirality, the produced GWs may be significantly red or blue tilted and non-Gaussian. One of the well known models of inflation, where non-Abelian gauge fields generate accelerated expansion, is the Gauge-flation model that was originally proposed in Refs.~\cite{Maleknejad:2011jw, Maleknejad:2011sq}.  Gauge-flation is related to  Chromo-Natural inflation \citep{Adshead:2012kp} that contains an axion coupled to $SU(2)$ gauge fields.  Gauge-flation can be formally obtained 
from chromo-natural inflation after integrating out an axion field near the minimum of the axion potential \citep{Adshead:2012qe, SheikhJabbari:2012qf, Maleknejad:2012dt, Maleknejad:2012fw}. 
The original formulation of both models is ruled out by {\it Planck} observations
\citep{Namba:2013kia, Dimastrogiovanni:2012ew, Adshead:2013nka, Adshead:2013qp}. However, both models can  be made consistent with current CMB bounds if
 the gauge symmetry is spontaneously broken by a Higgs sector \citep{Adshead:2017hnc, Adshead:2016omu}. Interestingly Higgsed gauge-flation and Higgsed Chromo-natural inflation give somewhat different predictions for the shape of the produced GW spectrum.
  
Recent  interest in potentially distinguishable signatures from the standard vacuum fluctuations by future B-mode experiments, like LiteBIRD, has led to a number of  generalizations of  gauge-field-driven GW  models. In particular considering a spectator axion sector coupled to non-abelian gauge fields has significantly opened up the parameter space of these models \cite{Dimastrogiovanni:2016fuu, Agrawal:2018mrg, Watanabe:2020ctz, Mirzagholi:2020irt, Fujita:2017jwq}.
 It was recently demonstrated  \cite{Thorne:2017jft} that Chromo-Natural inflation as a spectator sector for the scalar single-field inflation can be in agreement with the current data, while at the same time generating potentially distinguishable observable signatures for the tensor modes. In Ref.~\cite{Fujita:2018ndp} it was shown that in spectator Chromo-Natural inflation, depending on the choice of the axion potential, all three possible tensor tilts may be generated: flat, red and blue. In addition to that, peaked or oscillating GW spectra are also possible for well-motivated axion potentials.
 Since in Gauge-flation  there is much less freedom due to the absence of the axion field, a question arises: what are the possible GW spectra arising from a spectator Gauge-flation sector?

In this work we demonstrate the viability of the spectator Gauge-flation scenario, study its predictions and limitations and also provide a comparison with predictions of related models. The paper is organised as follows: In Section \ref{Sec:Framework} we introduce the framework for non-Abelian gauge field inflation and then embed it as a spectator sector for  scalar single-field inflation. In Section \ref{Sec:Viability} we discuss the necessary conditions  for the $SU(2)$ sector to be subdominant, as compared to the inflaton sector. This ensures that the scalar fluctuations will be dominated by the inflaton sector and can be made to  agree with the observational constraints, for example by considering an $\alpha$-attractor inflationary potential.
Keeping the non-Abelian sector subdominant leads to  an upper bound for the amplitude enhancement of the tensor power spectra. In Section \ref{Sec:Tilt} we discuss predictions for the primordial tensor tilt and its dependence on the parameters of the theory. 
We use a well-known $\alpha$-attractor model as the inflaton sector, since it can provide an arbitrarily low amount of vacuum-generated GWs (at least in principle), while at the same time obeying the constraints for the scalar fluctuations.
We conclude in Section \ref{Sec:Summary}.

\section{Framework}\label{Sec:Framework}
\subsection{The model}
In this section we describe the theory of Gauge-flation and its embedding as a spectator sector for inflation. 
The Gauge-flation action is given by \cite{Maleknejad:2011jw, Maleknejad:2011sq}
\begin{equation}
S=\int d^4 x \sqrt{-\text{det}(g_{\mu\nu})}\left[ \dfrac{M_{\text{Pl}}^2}{2}R 
-\dfrac{1}{4}F^a_{\mu\nu}F^{a\, \mu\nu}+\dfrac{\kappa}{96}\left(F^a_{\mu\nu}\tilde{F}^{a\, \mu\nu}\right)^2 \right],
\end{equation}
where $R$ is the space-time Ricci scalar, 
$F^a_{\mu\nu}=\partial_{\mu} A^a_{\nu}-\partial_{\nu} A^a_{\mu}-g \epsilon^{abc}A^b_{\mu}A^c_{\nu}$ is the field strength of an $SU(2)$ gauge field $A^a_{\mu}$, $\tilde{F}^{a\, \mu\nu}=\epsilon^{\mu\nu\rho\sigma}F^a_{\rho\sigma}/\left(2\sqrt{-\text{det}(g_{\mu\nu})}\, \right)$ its dual (where $\epsilon^{\mu \nu\alpha\beta}$ is the antisymmetric tensor and $\epsilon^{0123}=1$), { $\kappa>0$ is a parameter measured in units of $M_{\rm pl}^{-4}$} and $g$ is the gauge field coupling.

We will work with the FLRW metric
\begin{equation}
ds^2=-dt^2+a(t)^2 \delta_{ij}dx^i dx^j,
\end{equation}
where $i,j$ indicate the spatial directions. An isotropic solution for the background is given by the following configuration of the gauge field
\begin{gather}
A^a_0=0,\\
A^a_i=\delta ^a_i a(t)Q(t) \label{ansatzA}
\end{gather}
and it has been shown to be an attractor solution \cite{Maleknejad:2011sq}.
For this ansatz the closed system of equations for the vacuum expectation value (VEV) of the gauge field $Q(t)$ and the Hubble parameter $H(t)$ is given by
\begin{gather}
M_{\text{Pl}}^2\dot{H}=-\left( ( \dot{Q}+H Q)^2+ g^2 Q^4\right) \\
M_{\text{Pl}}^2 H^2=\dfrac{1}{2}\left( ( \dot{Q}+H Q)^2+g^2 Q^4+\kappa g^2 Q^4 ( \dot{Q}+H Q)^2\right), \\
\left(1+\kappa g^2 Q^4\right)\left( \ddot{Q}+3H\dot{Q}+\dot{H}Q\right)+2g^2Q^3\left(1+\kappa \dot{Q}^2\right)+2H^2Q=0,
\end{gather}
where an overdot  denotes a derivative with respect to cosmic time $t$.
 
We now introduce a scalar field $\varphi(t)$ with a potential $V(\varphi)$ that is responsible for driving inflation and consider the Gauge-flation terms as a spectator sector, i.e.
\begin{equation}\label{spectatorAction}
S=\int d^4 x \sqrt{-\text{det}(g_{\mu\nu})}\left[ \dfrac{M_{\text{Pl}}^2}{2}R -\dfrac{1}{2}(\partial \varphi)^2-V(\varphi)
-\dfrac{1}{4}F^a_{\mu\nu}F^{a\, \mu\nu}+\dfrac{\kappa}{96}\left(F^a_{\mu\nu}\tilde{F}^{a\, \mu\nu}\right)^2 \right].
\end{equation}
Up to gravitational interactions the dynamics of the inflaton sector is completely decoupled from the dynamics of the gauge field. This allows the inflaton field $\varphi(t)$  to be responsible for the predictions for scalar fluctuations.  At the same time the gravitational waves generated by the gauge field sector can lead to observable signatures in the tensor power spectra. In this paper we will not consider scalar fluctuations and refer to Ref.~\cite{Dimastrogiovanni:2016fuu} where scalar fluctuations were studied for a related model, where the spectator sector involved an axion coupled to an $SU(2)$ field  through a Chern-Simons term (which we call spectator Chromo-natural inflation)\footnote{\textcolor{black}{
Contrary to Chromo-natural inflation, scalar metric fluctuations in Gauge-flation must be taken into account at the linear level. 
In the context of spectator Gauge-flation, this could couple the inflaton fluctuations to the Gauge-flation sector, albeit through slow-roll parameters.
If one chooses to ignore them, the effect of the scalar fluctuations of the Gauge-flation sector are transferred to the inflaton sector through slow-roll suppressed gravitational couplings, making them  heavily suppressed for  parameters that are relevant for spectator  models. 
Properly taking metric fluctuations into account could provide interesting phenomenology, especially given the recent work on non-linear effects \cite{Papageorgiou:2018rfx,Papageorgiou:2019ecb}. This would deviate significantly from the context of our current analysis
and is left for future work.
}}. 
We will thus focus our attention solely on the tensor sector, leading to the production of GW's, \textcolor{black}{taking the scalar power spectrum to be dominated by the inflaton sector}.

Using the ansatz of Eq.~\eqref{ansatzA} the background system of equations in the presence of the inflaton field changes to
\begin{gather}
M_{\text{Pl}}^2\dot{H}=-\left( ( \dot{Q}+H Q)^2+ g^2 Q^4\right)-\frac{1}{2}\dot{\varphi}^2,\label{Hdot} \\
M_{\text{Pl}}^2 H^2=\dfrac{1}{3}\left( \dfrac{1}{2}\dot{\varphi}^2+V(\varphi)\right)+\dfrac{1}{2}\left( ( \dot{Q}+H Q)^2+g^2 Q^4+\kappa g^2 Q^4 ( \dot{Q}+H Q)^2\right), \label{HSquare}\\
\left(1+\kappa g^2 Q^4\right)\left( \ddot{Q}+3H\dot{Q}+\dot{H}Q\right)+2g^2Q^3\left(1+\kappa \dot{Q}^2\right)+2H^2Q=0,\\
\ddot{\varphi}+3H\dot{\varphi}+V_{\varphi}(\varphi)=0,\label{eomPhi}
\end{gather}
where $V_{\varphi}(\varphi)=\partial V(\varphi)/\partial \varphi $.
The standard Hubble slow roll parameters are defined as
\begin{gather}\label{slow-roll}
\epsilon=-\frac{\dot{H}}{H^2}, \quad \eta=-\frac{\ddot{H}}{2H\dot{H}}=\epsilon-\frac{\dot{\epsilon}}{2\epsilon H}.
\end{gather}
{The slow-roll parameter $\epsilon$ contains contributions from} the scalar (inflaton) and the gauge field (spectator) sectors. The various contributions can be written as
\begin{gather}
\epsilon=\epsilon_{\varphi}+\epsilon_{Q_E}+\epsilon_{Q_B}, 
\end{gather}
where
\begin{gather}\label{epsilon_def}
\epsilon_{\varphi}=\frac{\dot{\varphi}^2}{2M_{\text{Pl}}^2 H^2},\quad \epsilon_{Q_E}=\frac{( \dot{Q}+H Q)^2}{M_{\text{Pl}}^2 H^2},\quad \epsilon_{Q_B}=\frac{g^2Q^4}{M_{\text{Pl}}^2 H^2}. 
\end{gather}
 Throughout this work we  assume -- and check -- that the inflaton field $\varphi(t)$ dominates the energy budget of the theory. \textcolor{black}{Furthermore, we require  that the evolution of the inflaton sector
 controls the evolution of the Hubble scale,}
  $\epsilon_{\varphi}\gg \epsilon_Q$, where $\epsilon_Q=\epsilon_{Q_E}+\epsilon_{Q_B}$, and hence $\epsilon \simeq \epsilon_{\varphi}$. 
Despite this regime of interest, we  keep the analytic part of our analysis as general as possible and clearly state the approximations wherever they are necessary for making analytical progress.

Although the inflationary era is dominated by $\varphi(t)$, in the same way as in the original Gauge-flation approach we will assume that the gauge field also slow-rolls together with the inflaton field\footnote{A fast-rolling spectator gauge-flation sector can also lead to GW production. However, some fine-tuning is required to bring it in the observable window. We thus do not pursue this regime further.}. Hence for the later analysis we define
\begin{gather}\label{delta}
\delta=-\frac{\dot{Q}}{HQ}, \quad \gamma=\frac{g^2Q^2}{H^2}.
\end{gather}
We   require $\delta\ll 1$ to ensure that the gauge field slow-rolls long enough, to secure the needed amount of e-folds for inflation. The parameter $\gamma$  is a characteristic quantity of the model. It was shown in  Ref.~\cite{Namba:2013kia} that for $\gamma < 2$ the scalar perturbations are tachyonically unstable. We thus restrict our analysis to the stable region with $\gamma > 2$. For the tensor sector this parameter characterises the enhancement of one of the polarizations for the tensor perturbation with respect to the gravitational wave background coming from the inflaton sector. So far there were no theoretical upper bounds on this parameter, 
only the observational constraints coming from the tensor-to-scalar ratio $r$. As we will see in the next subsection, for  spectator Gauge-flation there exists an upper bound $\gamma_{\text{max}}$ which is determined solely from the self-consistency of the theory and the slow-roll conidtions. 
For a given set of parameters $g, \epsilon$ and $H$, the upper bound on $\gamma$ allows for an estimation of the 
 maximal enhancement for the tensor power spectra and thus a theoretical upper bound on $r$. 

\subsection{Background parameters}\label{Sec:BackgroundParameters}

In this subsection we  collect all the expressions for the background parameters that are relevant for the tensor power spectra computation. To start with, there are two equivalent ways to write down the first slow-roll parameter in terms of background quantities. The first one follows directly from Eqs.~\eqref{Hdot} and \eqref{slow-roll}, i.e. 
\begin{gather}
\epsilon =\frac{1}{M_{\text{Pl}}^2}Q^2\left((1-\delta)^2+\gamma\right)+\epsilon_{\varphi}. \label{epsilon1}
\end{gather}
Another way is to use the combination $\dot{H}+ 2H^2$, which  through Eqs.~\eqref{Hdot} and~\eqref{HSquare} leads to
\begin{gather}
\epsilon=2-\frac{\kappa g^2Q^6}{M_{\text{Pl}}^2}\left(1-\delta\right)^2
+\frac{1}{3}\epsilon_{\varphi}-\Upsilon\label{epsilon2} \, .
\end{gather}
We  define
\begin{gather}\Upsilon=\frac{2V}{3M_{\text{Pl}}^2 H^2} \, ,
\label{Upsilon}
\end{gather}
 which is determined by the scalar field inflaton potential and can be taken to be approximately constant for plateau models of inflation.
 From Eqs.~\eqref{slow-roll} and \eqref{epsilon1} we derive the second slow-roll parameter
\begin{gather}
\eta
=\frac{Q^2}{M_{\text{Pl}}^2}\left((1-\delta)^2+(1-\delta)\frac{\dot{\delta}}{\epsilon H}+\gamma\frac{\delta}{\epsilon}\right)+\delta-
\frac{\epsilon_{\varphi}}{\epsilon}\left(\delta-\eta_{\varphi}\right), \label{eta1}
\end{gather}
with $\eta_{\varphi}=-\frac{\ddot{\varphi}}{H\dot{\varphi}}$. An alternative derivation follows from Eqs.~\eqref{slow-roll} and \eqref{epsilon2}
\begin{gather}
\eta=\epsilon-(2-\Upsilon-(\epsilon-\frac{1}{3} \epsilon_{\varphi}))\left(\frac{\dot{\delta}}{H \epsilon(1-\delta)}+\frac{3\delta}{\epsilon}\right)-\frac{\epsilon_{\varphi}}{3}\left(1-\frac{\eta_{\varphi}}{\epsilon}\right)+\frac{2}{3}\left(\frac{\epsilon_{\varphi} V_{\varphi}}{\dot{\varphi}H\epsilon}+\frac{V}{H^2 M_{\text{Pl}}^2 }\right). \label{eta2}
\end{gather}

{Up to this point, the above equations are exact. If the inflaton is assumed to dominate} the total energy budget, Eq.~\eqref{eta1} leads to $\eta\simeq \frac{\epsilon_{\varphi}}{\epsilon}\eta_{\varphi}$.
Substituting that into Eq.~\eqref{eta2} and neglecting\footnote{The arguments for neglecting this term are discussed in Ref.~\cite{Maleknejad:2011sq}. We  numerically checked the validity of this approximation} $\frac{\dot{\delta}}{H (1-\delta)}$,  we find
\begin{gather}
 \delta\simeq
\frac{\epsilon}{3(2-\Upsilon-(\epsilon-\frac{1}{3}\epsilon_{\varphi}))}\left(\epsilon-\frac{2}{3}\frac{\epsilon_{\varphi}}{\epsilon}\eta_{\varphi}-\frac{\epsilon_{\varphi}}{3}+\Upsilon\left(1-\frac{\epsilon_{\varphi}}{\epsilon}\right)\right), \label{deltasimple}
\end{gather}
where we have used $\epsilon_V=\frac{M_{\text{Pl}}^2}{2}\left( \frac{V_{\varphi}}{V}\right)^2 \simeq \epsilon_{\varphi}  $ and $\dot{\varphi}=-HM_{\text{Pl}}\sqrt{2\epsilon_{\varphi}}$ is chosen to be negative without  loss of generality.

In addition to that,   Eqs.~\eqref{delta} and \eqref{epsilon2} lead to \begin{gather}\label{kappa}
\kappa
=\frac{1}{H^2\gamma Q^2}\frac{(1-\delta)^2+\gamma}{(1-\delta)^2}\frac{2-\left(\epsilon_Q+\frac{2}{3}\epsilon_{\varphi}+\Upsilon\right)
}{\epsilon_Q}.
\end{gather}
Moreover, using Eq.~\eqref{epsilon1} we derive a relation that allow us to eliminate $M_{\text{Pl}}$ from the equations
\begin{equation}\label{Mpl}
M_{\text{Pl}}= Q\sqrt{\frac{(1-\delta)^2+\gamma}{\epsilon_Q}} \, ,
\end{equation}
where $\epsilon_Q$ is the first slow-roll parameter for the gauge field sector, i.e. $\epsilon_Q=\epsilon_{Q_E}+\epsilon_{Q_B}$, as defined in Eq.~\eqref{epsilon_def}.
The relations above with the inflaton sector set to zero coincide with relations obtained in Ref.~\cite{Namba:2013kia} for the pure Gauge-flation model, which provides a consistency check for our analysis. Eq.~\eqref{deltasimple} was derived strictly under the assumption of a dominant inflaton sector and thus does not reduce to the gauge-flation result for $\epsilon_\phi\to 0$, which is not true for all equations before it, which are exact and hence applicable to any gauge-flation scenario, with or without a separate inflaton sector.

Finally, requiring $\delta\ll 1$ 
 and using Eqs.~\eqref{delta} and \eqref{epsilon1} leads to
\begin{equation}\label{euqationForGammamax}
1-\frac{\epsilon_{\varphi}}{\epsilon}\simeq\frac{H^2}{M_{\text{Pl}}^2 g^2 \epsilon}\gamma(\gamma+1).
\end{equation}
\textcolor{black}{From Eq.~\eqref{euqationForGammamax} we obtain 
\begin{equation}\label{trueGammaMax}
\gamma \simeq-\frac{1}{2}+\frac{1}{2}\sqrt{1+4M_{\text{Pl}}^2\frac{g^2 \epsilon}{H^2}\left(1-\frac{\epsilon_{\varphi}}{\epsilon}\right)},
\end{equation}
\begin{figure}
\centering
  \includegraphics[width=0.45\textwidth]{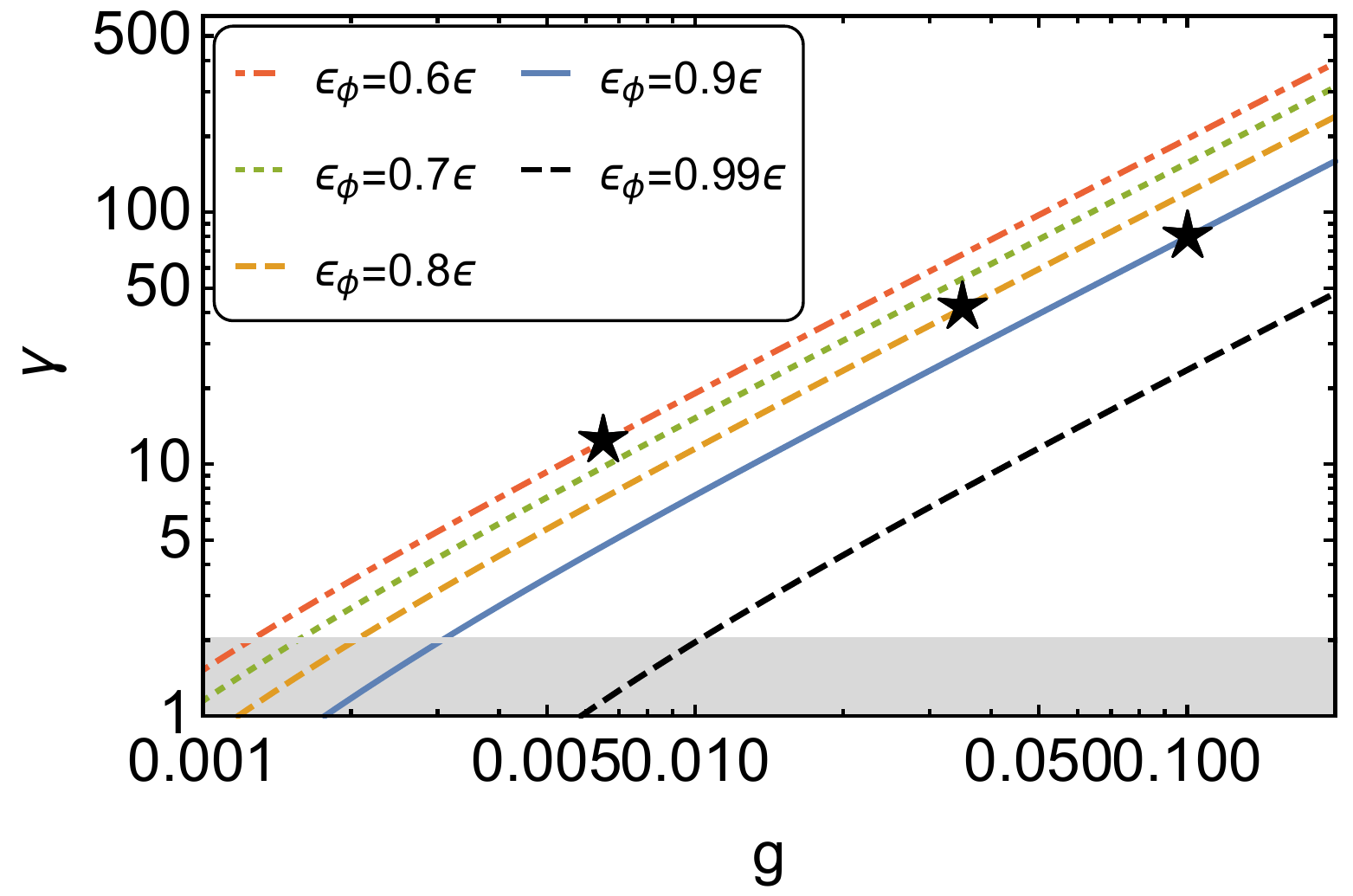}
\caption{\textcolor{black}{
 The dependence of $\gamma$ on the gauge coupling $g$ for  $\epsilon_{\varphi}/ \epsilon=0.6, 0.7, 0.8, 0.9, 0.99$  (red-dot-dashed, green-dotted, orange-dashed, blue-solid and black-dashed lines respectively) for a fixed scalar sector with $\epsilon_{\varphi}=2\times 10^{-4}$ and $H=6 \times 10^{-6} M_{\rm pl}$. The shaded grey area shows the excluded region $\gamma <\sqrt{2}$, dictated by stability of scalar perturbations. Stars represent the  theoretically maximum allowed value of $\gamma$ for fixed $g$ and $\epsilon_{\varphi}/ \epsilon$, in particular $\lbrace\gamma=12, g=0.0065 \rbrace, \lbrace\gamma=41.5, g=0.035 \rbrace, \lbrace\gamma=79.5, g=0.1 \rbrace$,  corresponding to  $\epsilon_{\varphi}/ \epsilon=0.6, 0.8, 0.9$ respectively.
 We use these values in our numerical simulations in  subsequent sections, unless stated otherwise.
  }
}
 \label{fig:gammaMax}
\end{figure}
which is the maximally allowed value of $\gamma$ in the \textit{spectator Gauge-flation} model for  given values of $H, \epsilon, \epsilon_{\varphi}$ and a gauge coupling $g$. The result of Eq.~\eqref{trueGammaMax} is obtained without the requirement  $\epsilon \simeq \epsilon_{\varphi}$ and is rather generic.
One can see that the parameter $\gamma$ cannot be chosen arbitrarily high any more, but reaches its maximal value given by Eq.~\eqref{trueGammaMax} due to the restrictions of the theory. 
The maximum value of $\gamma$ is achieved when the energy budget is completely controlled by the gauge sector, i.e. when $\frac{\epsilon_{\varphi}}{\epsilon}\ll 1$, meaning that $\epsilon$ is dominated by $\epsilon_Q$.  In the spectator case $\epsilon\simeq\epsilon_{\varphi}$ and $\gamma$ is limited to a smaller range of values, with a magnitude that depends only on $g$ (with fixed $H$ and $\epsilon$).
The dependence of $\gamma$ on the gauge coupling for different $\epsilon_{\varphi}/ \epsilon$ is shown in  Fig.~\eqref{fig:gammaMax}. Simply from the stability condition of scalar perturbations\footnote{\textcolor{black}{For $\gamma<2$ scalar perturbations experience a tachyonic instability, see Ref.~\cite{Namba:2013kia} for a detailed discussion.}}  $\gamma>2$ we obtain the mimimum value for the gauge coupling in \textit{spectator Gauge-flation}
\begin{equation}\label{gmin}
g_{\text{min}}\simeq \frac{\sqrt{6}H}{M_{\text{Pl}}\sqrt{\epsilon}\,\sqrt{1-\frac{\epsilon}{\epsilon_{\varphi}}}}.
\end{equation}
We can estimate the value of $g_{\text{min}}$ by relating  $H$ and $\epsilon$ to the amplitude of the scalar power spectrum $P_\zeta=\frac{H^2}{8 \pi^2 M_{\text{Pl}}^2 \epsilon } \simeq 2.2 \times 10^{-9}$, since we assume that the scalar power spectrum is dominated by the fluctuations in the inflaton sector. This results in a lower bound for a gauge coupling $g_{\text{min}}\simeq \frac{4\pi \sqrt{3P_\zeta}}{\sqrt{1-\frac{\epsilon}{\epsilon_{\varphi}}}} \simeq 0.0016$ for $\epsilon_{\varphi}/ \epsilon=0.6$ and $g_{\text{min}} \simeq 0.0032$ for $\epsilon_{\varphi}/ \epsilon=0.9$. The gauge coupling $g$ is also constrained from above by some value $g_{\rm max}$. As was described in detail in Ref.~\cite{Maleknejad:2012fw}, the $\kappa$-term needs to dominate over the charged matter loop effects. In particular, the requirement is 
\begin{equation}\label{loopRequirement}
\kappa\gg \frac{3g_{\rm max}^4}{M_f^4}>\frac{3g_{\rm max}^4}{H^4},
\end{equation}
since for inflation the Compton wavelength that is relevant for loop effects has to be bigger than the Hubble horizon size, i.e. $1/M_f>1/H$. Here $M_f$ is a charged matter mass in a system. For the parameters used in this work, applicable for the $\alpha$-attractor model discussed in Sec.~\ref{Sec:Viability}, for the choice $\alpha= 0.01$ the maximum value of the gauge coupling\footnote{\textcolor{black}{We made sure that the strong inequality of Eq.~\eqref{loopRequirement} is satisfied by at least two orders of magnitude.}} is $g_{\rm max}\simeq 0.02$, for $\alpha= 0.1$ it increases to  $g_{\rm max}\simeq 0.06$ and for  $\alpha=1$ it is   $g_{\rm max}\simeq 0.2$. For smaller values of $\alpha$, $g_{\rm max}$ reaches smaller magnitudes\footnote{\textcolor{black}{However a small $\alpha$ gives a small amplification, hence we will restrict our consideration for values $0.01 \leq\alpha\leq 1$.}} . With the knowledge of $g_{\rm max}$ the parameter $\gamma_{\text{max}}$ may be estimated as $\gamma_{\text{max}}\simeq-\frac{1}{2}+\frac{1}{2}\sqrt{1+\frac{g_{\rm max}^2 }{2\pi^2 P_\zeta}(1-\frac{\epsilon_{\varphi}}{\epsilon})}$. For $g_{\rm max}=0.06$ and $\epsilon_{\varphi}/ \epsilon=0.6$ this results into $\gamma_{\text{max}}\simeq 91$, for $\epsilon_{\varphi}/ \epsilon=0.95$ it equals $\gamma_{\text{max}}\simeq 32$ and for $\epsilon_{\varphi}/ \epsilon=0.99$ it is $\gamma_{\text{max}}\simeq 14$. These are the maximum values of $\gamma$ that may be attained for the $\alpha$-attractor model with $\alpha=0.1$. A complementary method for computing the maximum allowed tensor amplification based on
the back-reaction of the produced spin-$2$ particles
was presented in Ref.~\cite{Maleknejad:2018nxz}.
}

\textcolor{black}{Since $0<\epsilon_{\varphi}<\epsilon$, the right hand side of Eq.~\eqref{euqationForGammamax} is in the range  $0<\frac{H^2}{M_{\text{Pl}}^2 g^2 \epsilon}\gamma(\gamma+1)<1$. Hence, one finds
$
\gamma_{\text{GF}}=-\frac{1}{2}+\frac{1}{2}\sqrt{1+4M_{\text{Pl}}^2\frac{ g^2 \epsilon}{H^2}} \, ,
$
which is the maximum value of the parameter $\gamma$ for the case of the gauge field domination over the scalar sector, i.e. when $\epsilon_{\varphi}/ \epsilon \ll 1$, that limits to the case of \textit{pure Gauge-flation}. Here the subscript ``GF" stands for ``Gauge-flation".
Similarly, from the stability condition of scalar perturbations $\gamma_{\text{GF}}>2$ one finds the minimum value for the case of\textit{ the gauge field domination}
$
g_{\text{min, GF}}>\sqrt{\frac{6}{M_{\text{Pl}}^2\,\epsilon}}H.
$
}

\section{Viability of spectator Gauge-flation}\label{Sec:Viability}

In this section we examine the viability of  spectator Gauge-flation. We  consider and discuss the most important dynamics, using the example of an $\alpha$-attractor potential for the inflaton field.
To ensure that the gauge sector of Eq.~\eqref{spectatorAction} is a spectator sector, the energy density of the gauge fields must be subdominant to that of the inflaton
\begin{gather} 
\rho_{\varphi}\gg \rho_{Q_E}, \, \rho_{Q_B}, \, \rho_{Q_{\kappa}}\label{rhoEpsilonInequalities1}\end{gather}
where the definitions for the energy densities are given as \cite{Maleknejad:2011jw}
\begin{eqnarray}
\rho_{\varphi}&=&\dfrac{1}{2}\dot{\varphi}^2+V(\varphi), \label{rho1}\\
\rho_{Q_E}&=&\dfrac{3}{2}( \dot{Q}+H Q)^2,\\
\rho_{Q_B}&=&\dfrac{3}{2}g^2 Q^4,\\
\rho_{Q_{\kappa}}&=&\dfrac{3}{2}\kappa g^2 Q^4( \dot{Q}+H Q)^2. \label{rho4}
\end{eqnarray}
A similar condition must hold for the first slow-roll quantity $\epsilon$
\begin{gather} 
\epsilon_{\varphi}\gg \epsilon_{Q_E,} \,  \epsilon_{Q_B}, \label{rhoEpsilonInequalities2}
\end{gather}
meaning that the Hubble evolution is dominated by the rolling of the inflaton field (see Eq.~\eqref{epsilon_def}).
The above inequalities can be re-cast as relations between the VEVs of the inflaton and gauge fields
\begin{align}
\dfrac{1}{2}\dot{\varphi}^2 \gg& ( \dot{Q}+H Q)^2, \label{condit1}\\
\dfrac{1}{2}\dot{\varphi}^2 \gg& g^2 Q^4, \label{condit2}\\
\dfrac{1}{2}\dot{\varphi}^2+V(\varphi) \gg& \frac{3}{2} ( \dot{Q}+H Q)^2 \label{condit3}\\
\dfrac{1}{2}\dot{\varphi}^2+V(\varphi) \gg& \frac{3}{2} g^2 Q^4, \label{condit4} \\
\dfrac{1}{2}\dot{\varphi}^2+V(\varphi) \gg& \frac{3}{2} \kappa g^2 Q^4 ( \dot{Q}+H Q)^2. \label{condit5}
\end{align}
Let us note that for  spectator Chromo-natural inflation precisely the same inequalities of Eq.~\eqref{condit3} and \eqref{condit4} should hold, in addition to an inequality for the axion field  $\chi(t)$, i.e. $\dfrac{1}{2}\dot{\varphi}^2+V(\varphi) \gg \dfrac{1}{2}\dot{\chi}^2+U(\chi) $ that replaces Eq.~\eqref{condit5} since the $\kappa$-term can be thought as the analogue of the axion potential in  Chromo-natural inflation.

From Eqs.~\eqref{condit1} -- \eqref{condit5} one may see that $\epsilon_{\varphi}\gg \epsilon_{Q_E,} \,  \epsilon_{Q_B}$ implies $\rho_{\varphi}\gg \rho_{Q_E}, \, \rho_{Q_B}$, as well as $ \rho_{\varphi}\gg \rho_{Q_{\kappa}}$, if  $\kappa$ is not too large. 
Hence, the inequalities for the energy densities given in Eqs.~\eqref{condit3} and \eqref{condit4} as well as Eq.~\eqref{condit5} hold automatically when the corresponding inequalities for the slow-roll parameter $\epsilon$ (Eqs.~\eqref{condit1} and \eqref{condit2}) are satisfied. Therefore we  focus on showing the allowed parameter ranges to satisfy $\epsilon_{\varphi}\gg \epsilon_{Q_E,} \,  \epsilon_{Q_B}$, i.e. Eqs.~\eqref{condit1} and \eqref{condit2}, and confirm our findings with numerical simulations.
\begin{figure}
\centering
 \includegraphics[width=0.42\textwidth]{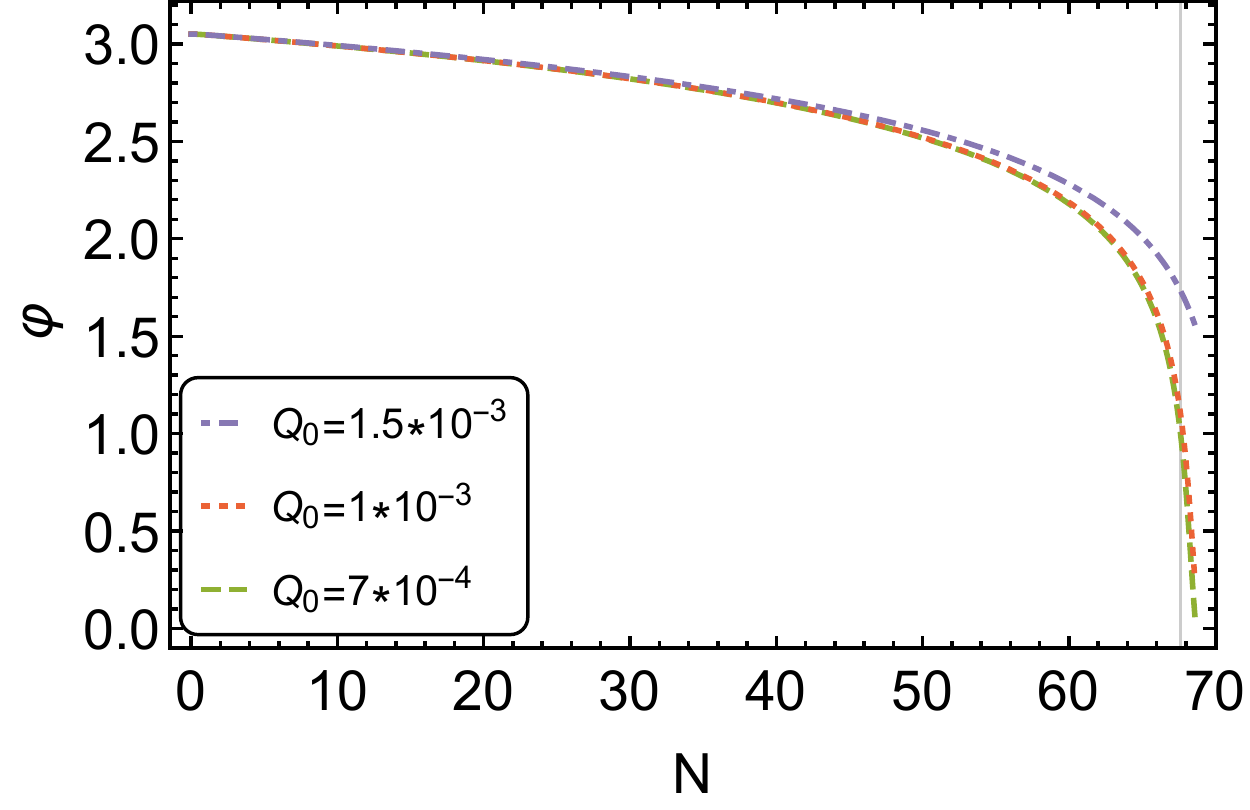}
\includegraphics[width=0.45\textwidth]{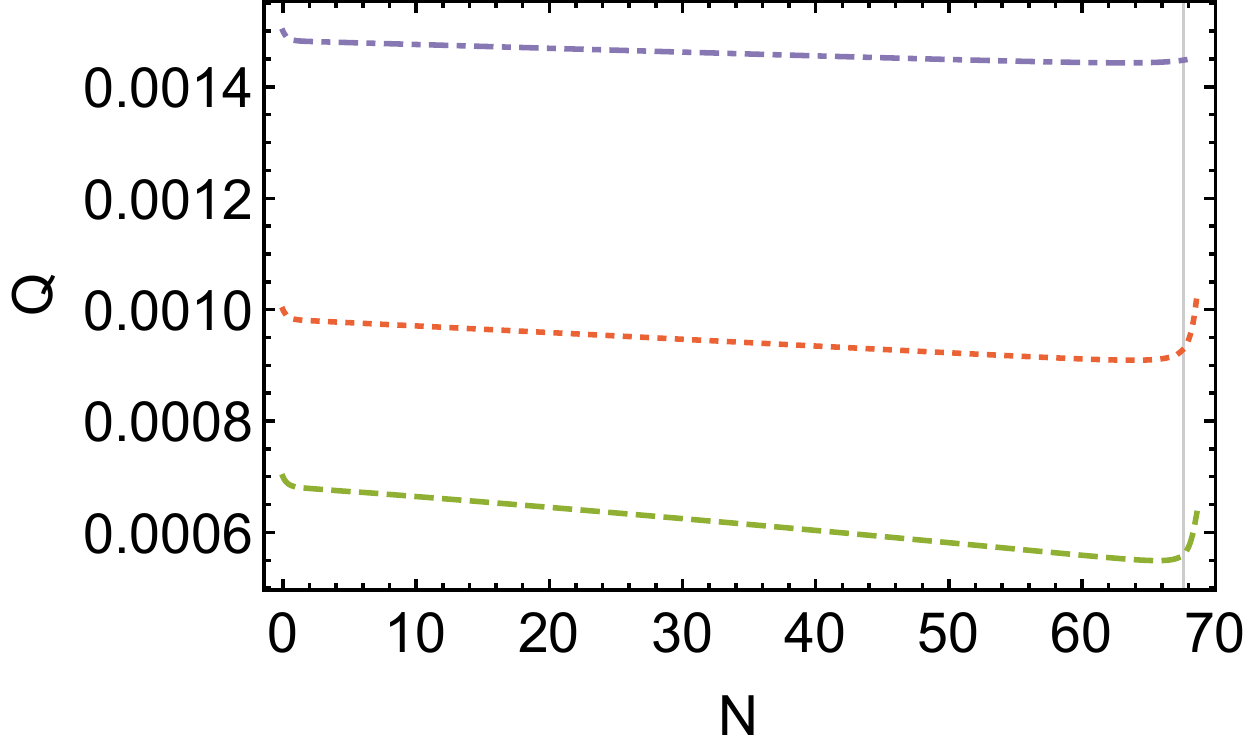}
\\
 \includegraphics[width=0.45\textwidth]{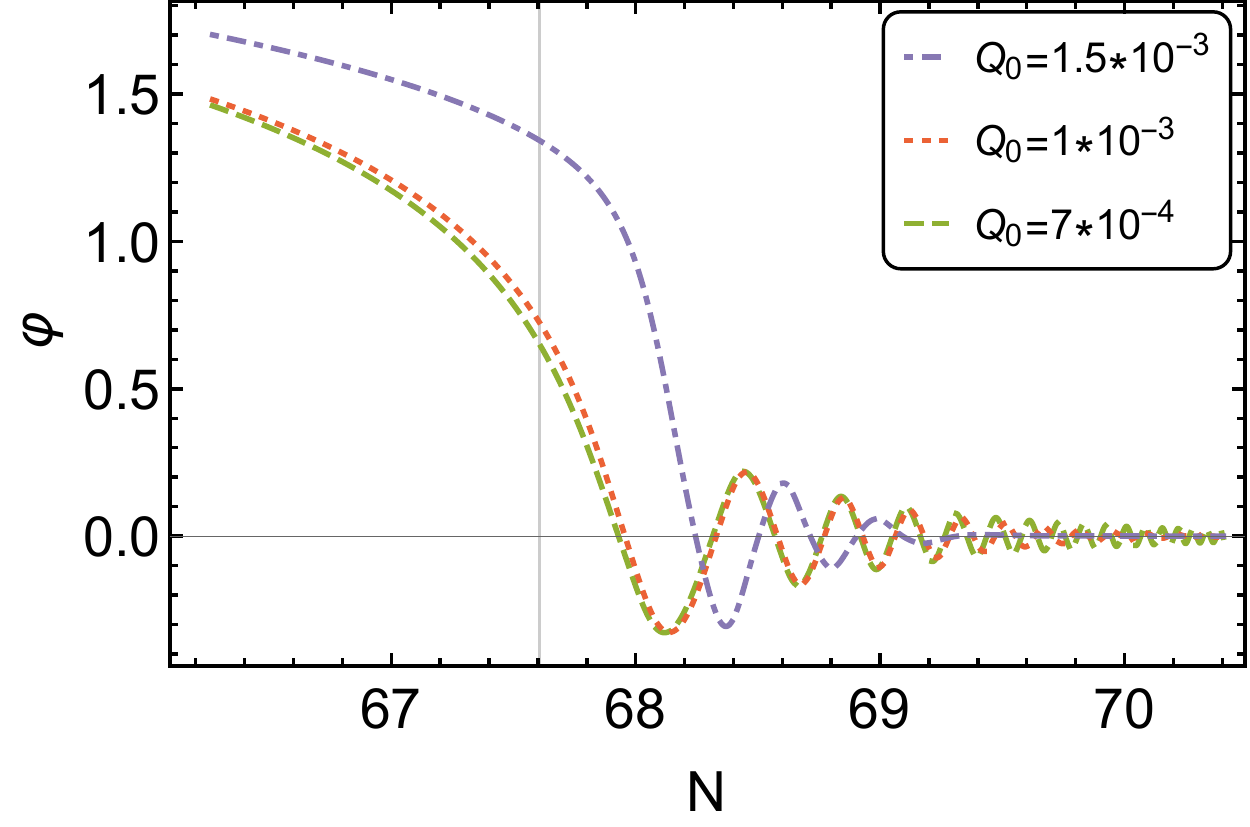}
 \includegraphics[width=0.45\textwidth]{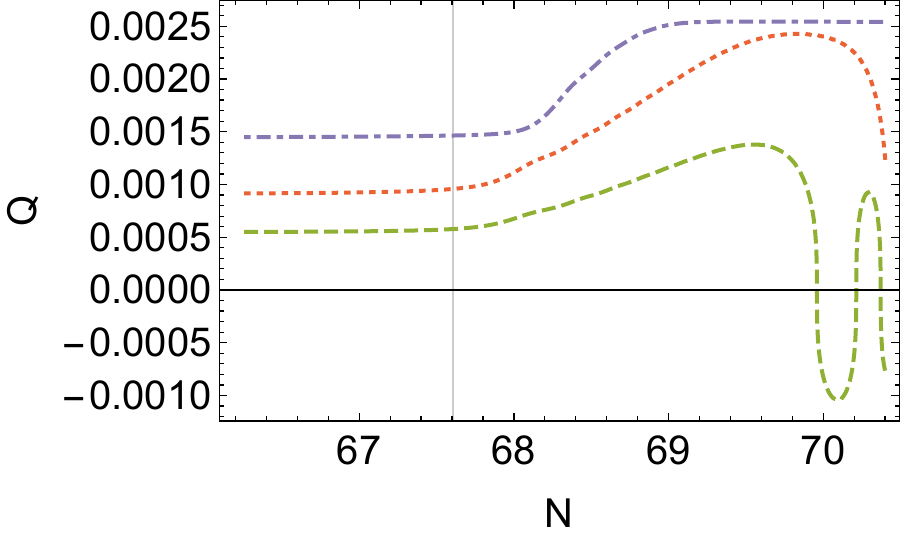}
\caption{
 {\it Upper left:} The dependence of  the inflaton field $\varphi$ on the $e$-folding number  $N$ for the  $\alpha$-attractor  T-model potential of Eq.~\eqref{Tpotential} for $Q_0/M_{\rm pl}=7\times10^{-4},
 1\times10^{-3}, 1.5\times10^{-3}$ (green-dashed, red-dotted and purple-dot-dashed lines respectively). The vertical grey grid line shows the end of inflation.
 {\it Upper right:} The dependence of  the  gauge field VEV $Q$ on the $e$-folding number  $N$ for the same potential and color coding. The solid grey grid line shows the end of inflation.
 {\it Lower Left:} The evolution of  the inflaton field $\varphi$ after the end of inflation for the same parameters and color-coding.
 {\it Lower right:} The post-inflationary evolution of the gauge field VEV $Q$ for the same parameters and color-coding.
 }
 \label{fig:PhiandQ}
\end{figure}

For illustrative purposes we  consider an $\alpha$-attractor model for the inflaton sector \cite{Ferrara:2013rsa, Ferrara:2013rsa2, Ferrara:2013rsa3, Kallosh:2013yoa, Kallosh:2013yoa2, Cecotti:2014ipa, Kallosh:2015lwa}. It is known that the universal predictions for the spectral index $n_s$ and tensor-to-scalar ratio $r$  are in  agreement with the latest {\it Planck} data
 \cite{Aghanim:2018eyx}. They are parametrised solely by the dimensionless coupling $\alpha$ and the number of e-folds $ N_*$ before the end of inflation when the CMB modes exit the horizon during inflation, i.e.
\begin{equation}
n_s=1-\frac{2}{N_*}, \quad r=\frac{12  \alpha}{N_*^2}.
\end{equation}
The $\alpha$-attractor T-model potential is given by
\begin{equation}
V(\phi)=\alpha \mu^2M_{\rm Pl}^2 \left((\tanh(\beta \phi/2))^2\right)^n, \label{Tpotential}
\end{equation}
where the parameters of the potential are chosen to be
\begin{figure}
\centering
 \includegraphics[width=0.45\textwidth]{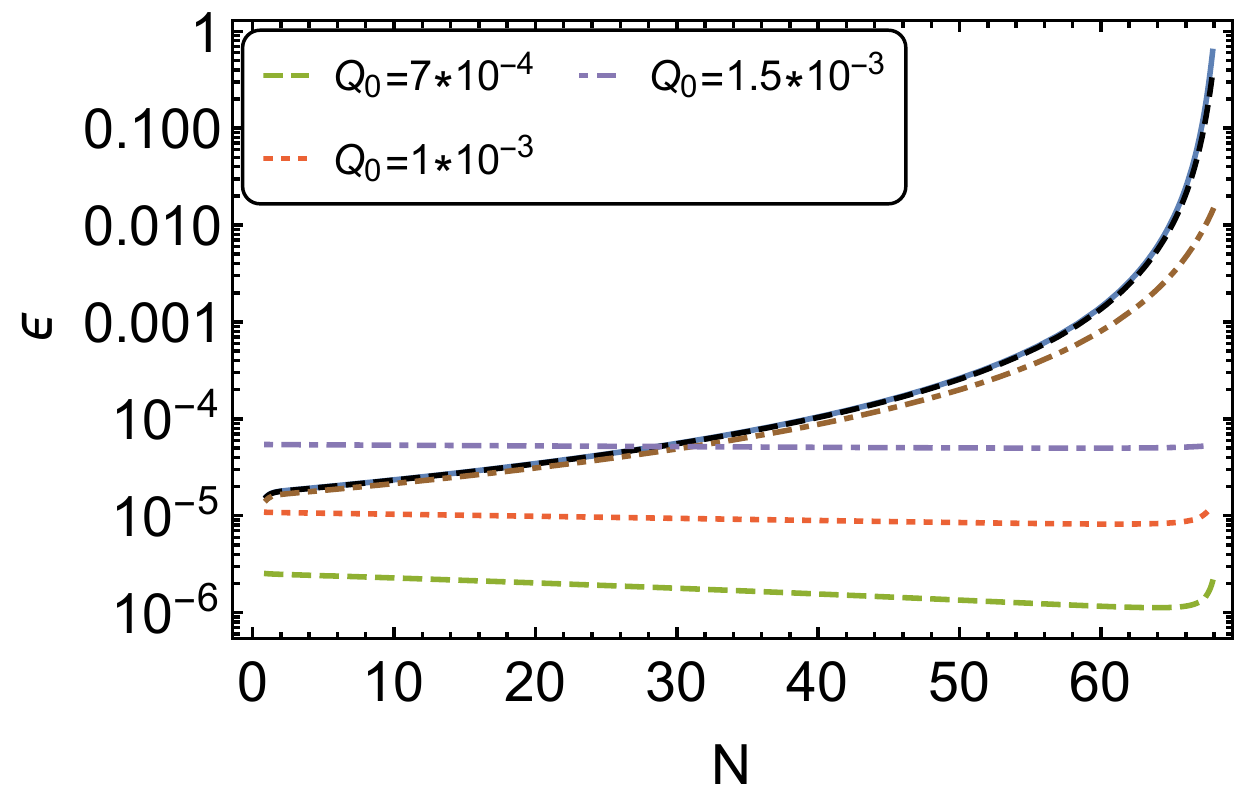}
\includegraphics[width=0.45\textwidth]{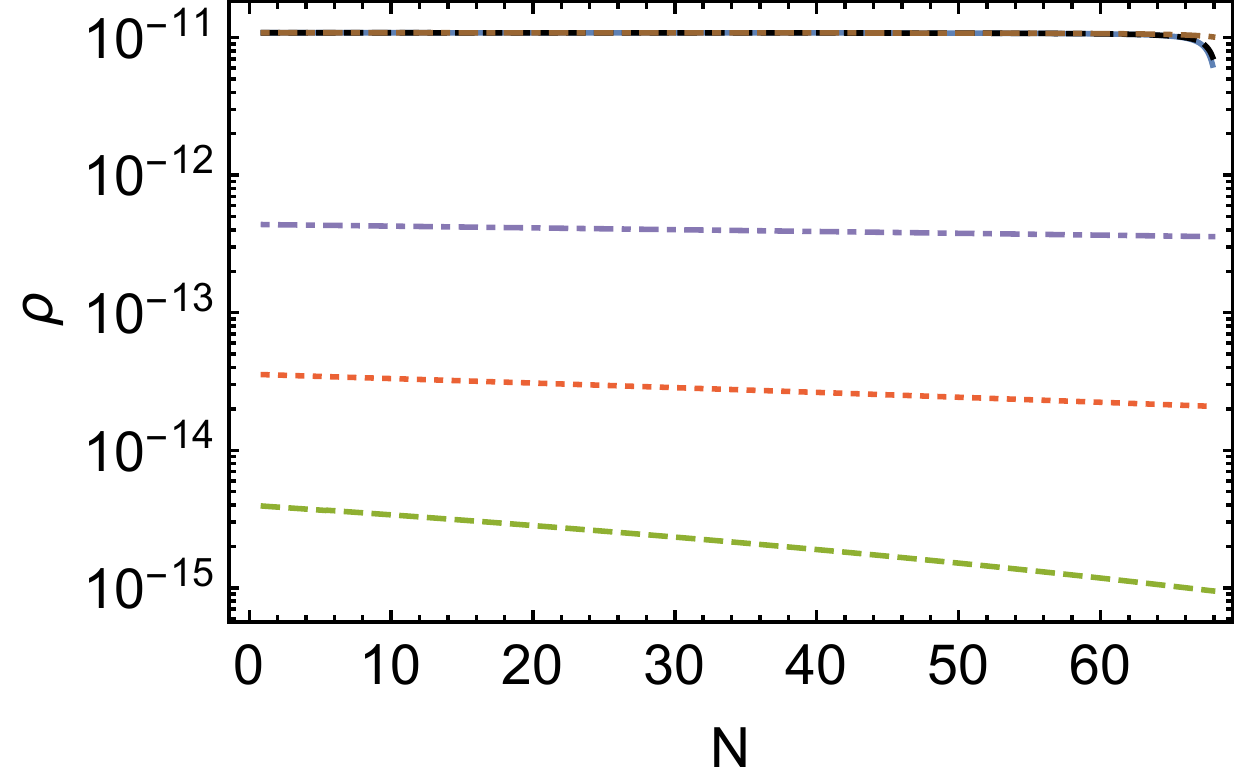}
\includegraphics[width=0.45\textwidth]{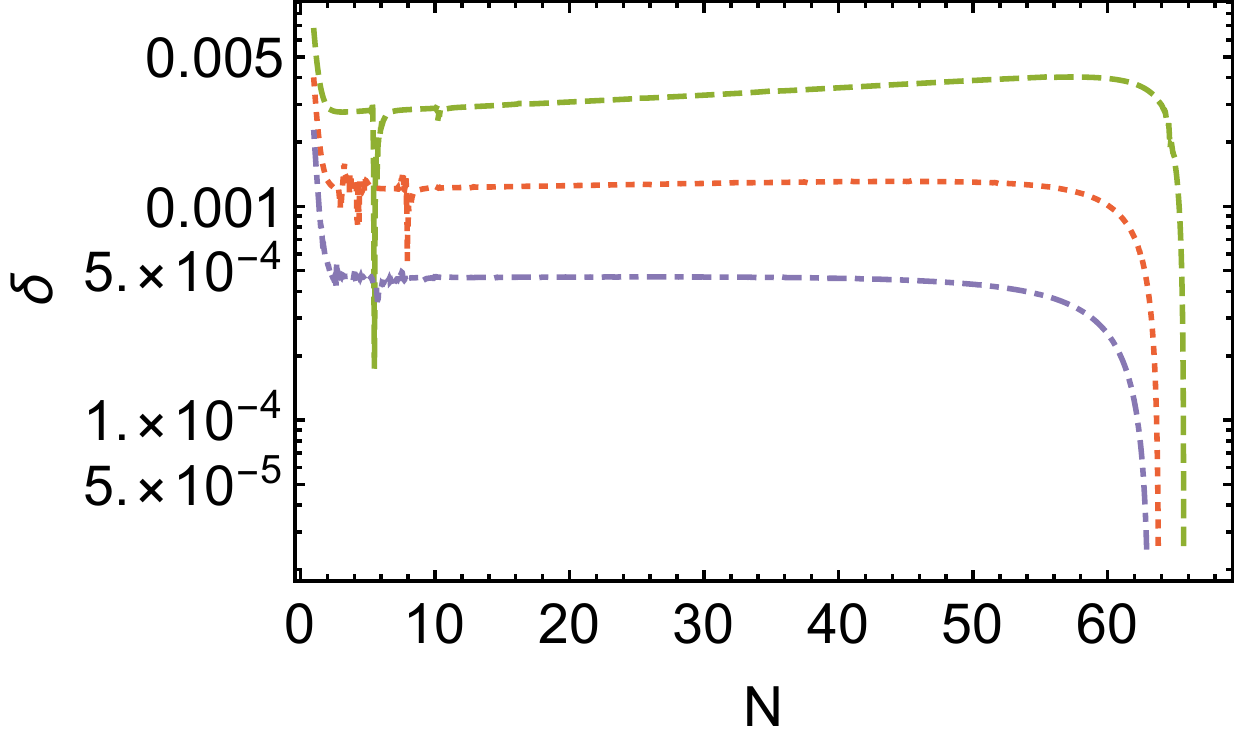}
 \includegraphics[width=0.45\textwidth]{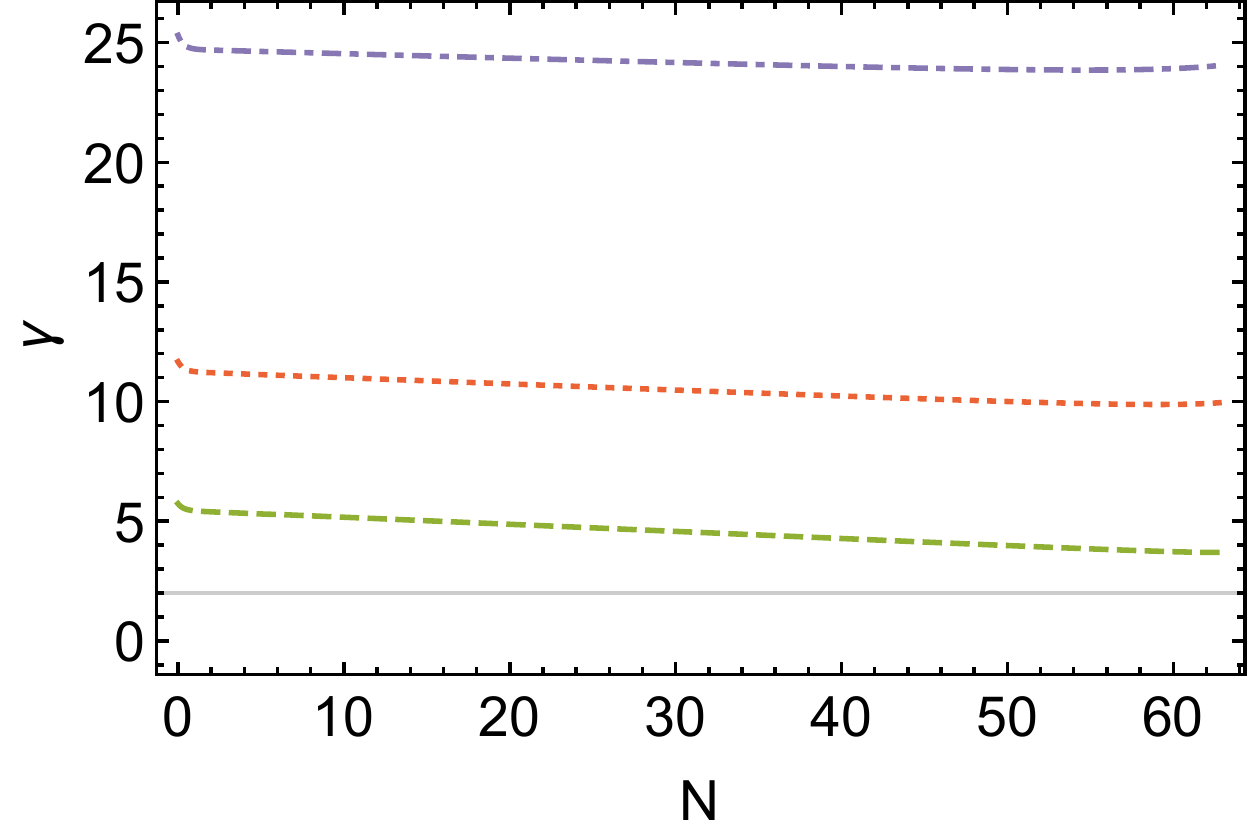}
\caption{
  {\it Top left:} Components $\epsilon_{Q_B}$ 
 as a function of the $e$-folding number $N$ 
for $Q_0/M_{\rm pl}=7\times10^{-4}, 1\times10^{-3}, 1.5\times10^{-3}$ (green-dashed, red-dotted and purple-dot-dashed lines respectively). 
   The blue-solid, black-dashed and brown-dot-dashed and  curved correspond to $\epsilon_{\varphi}$ for $Q_0/M_{\rm pl}=7\times10^{-4},
 1\times10^{-3}, 1.5\times10^{-3}$ respectively. \textcolor{black}{One can see that indeed $\gamma\simeq 12$ (which corresponds to the red-dotted line) is the maximal possible value for the  gauge sector to remain a spectator for the given set of parameters. For higher values of $\gamma$ the value of $\epsilon_{\varphi}$ will become smaller than $\epsilon_{Q_B}$.}
   {\it Top right:} Components $\rho_{\kappa}$ and their dependence on $N$ for $Q_0$ and color-coding. The very top curves  correspond to $\rho_{\varphi}$ and are practically indistinguishable.\\
{\it Bottom row:} The evolution of the parameter $\delta$ (left) and $\gamma$ (right) for the same parameters and color-coding. The solid grey grid line on the right panel shows the bound $\gamma=2$, below which scalar fluctuations in the theory are unstable. 
 }
 \label{fig:PlotsQ}
\end{figure}
\begin{equation}\label{paramsV}
\beta=\sqrt{2/3\alpha},\quad n=3/2,\quad \mu^2=1.08\times10^{-10}M_{\text{Pl}}^2, \quad \alpha=0.1
\end{equation}
and are used for numerical simulations in this section.  
In Section~\ref{Sec:Tilt} we examine the dependence of the results on the value of $\alpha$, while keeping the other parameters of the T-model potential fixed.
For the gauge sector we use\footnote{The naturalness of the $\kappa$-term and its domination over all the other dimension eight or higher contributions coming from gauge field or fermionic loops is discussed in Ref.~\cite{Maleknejad:2012fw}. Also note that $\kappa^{-1/4} > H_{\rm infl.}$.}
\begin{equation}\label{papametersKappa}
g=6.5\times10^{-3},\quad  \kappa=1.733\times10^{20} M_{\text{Pl}}^{-4},\quad \dot{Q}_0/ M_{\rm pl}^2=-10^{-10},\quad Q_0/ M_{\rm pl}=7\times10^{-4},
 10^{-3}, 1.5\times10^{-3},
\end{equation}
\textcolor{black}{where $Q_0, \dot{Q}_0$ is an initial value and initial velocity respectively for the gauge field VEV. }
 
\begin{figure}
\centering
 \includegraphics[width=0.45\textwidth]{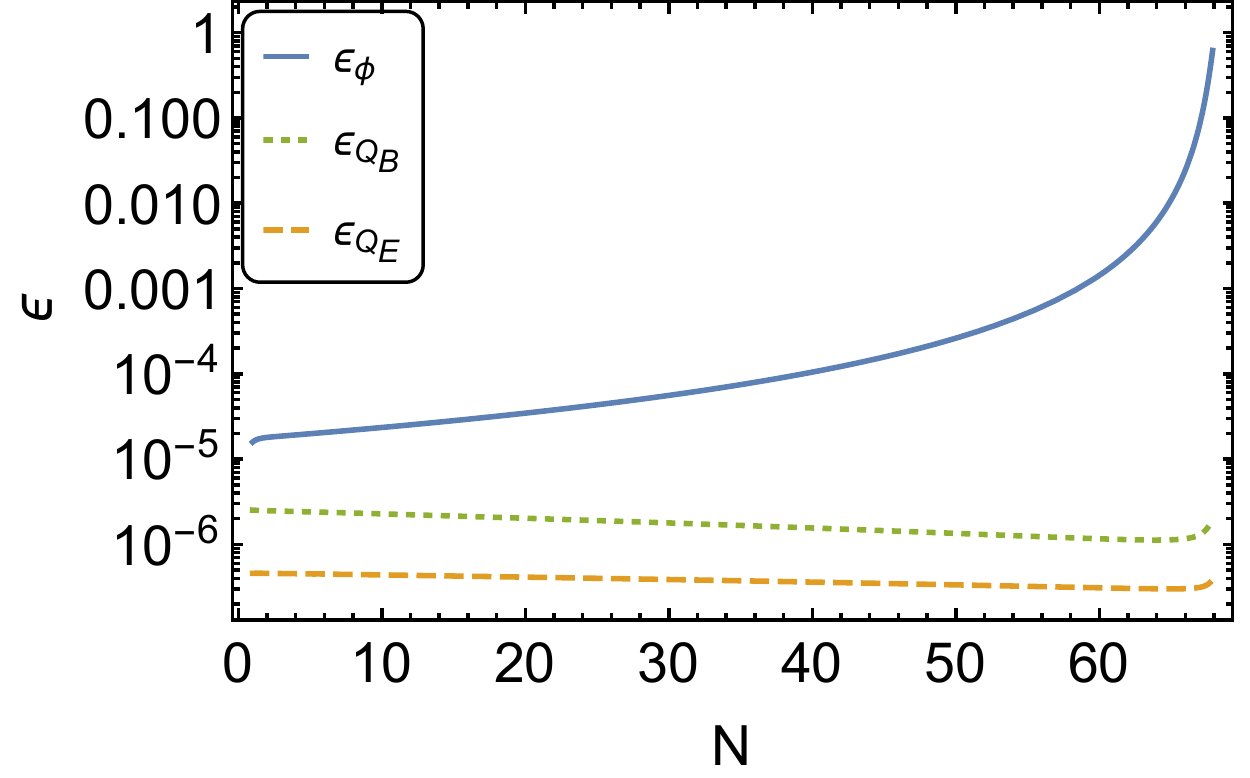}
\includegraphics[width=0.45\textwidth]{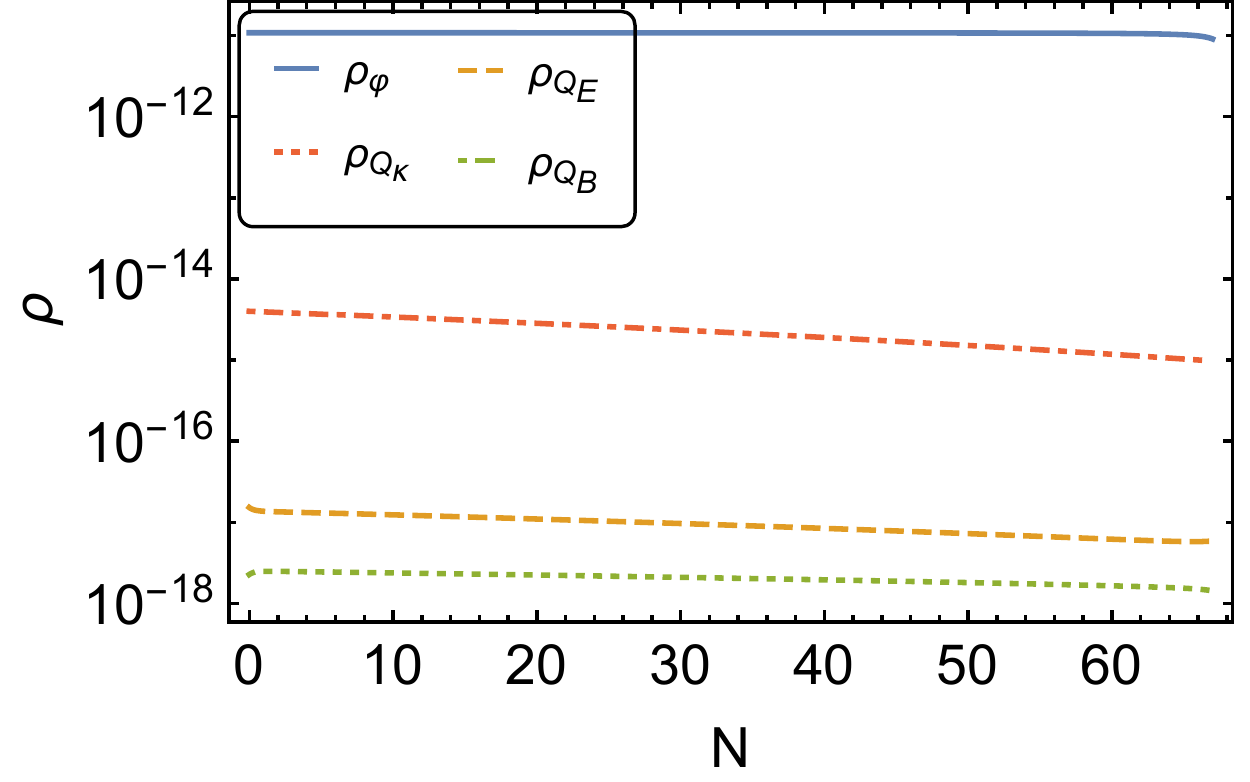}
\caption{ 
 {\it Left:} The evolution of components of the first slow-roll parameter $\epsilon$ with the number of e-folds $N$ for $Q_0/M_{\rm pl}=7\times10^{-4}$.
 {\it Right:} The evolution of components of the energy-density $\rho$ with the number of e-folds $N$ for the same parameters.\\
 }
 \label{fig:EpsilonRho}
\end{figure}

\begin{figure}
\centering
 \includegraphics[width=0.42\textwidth]{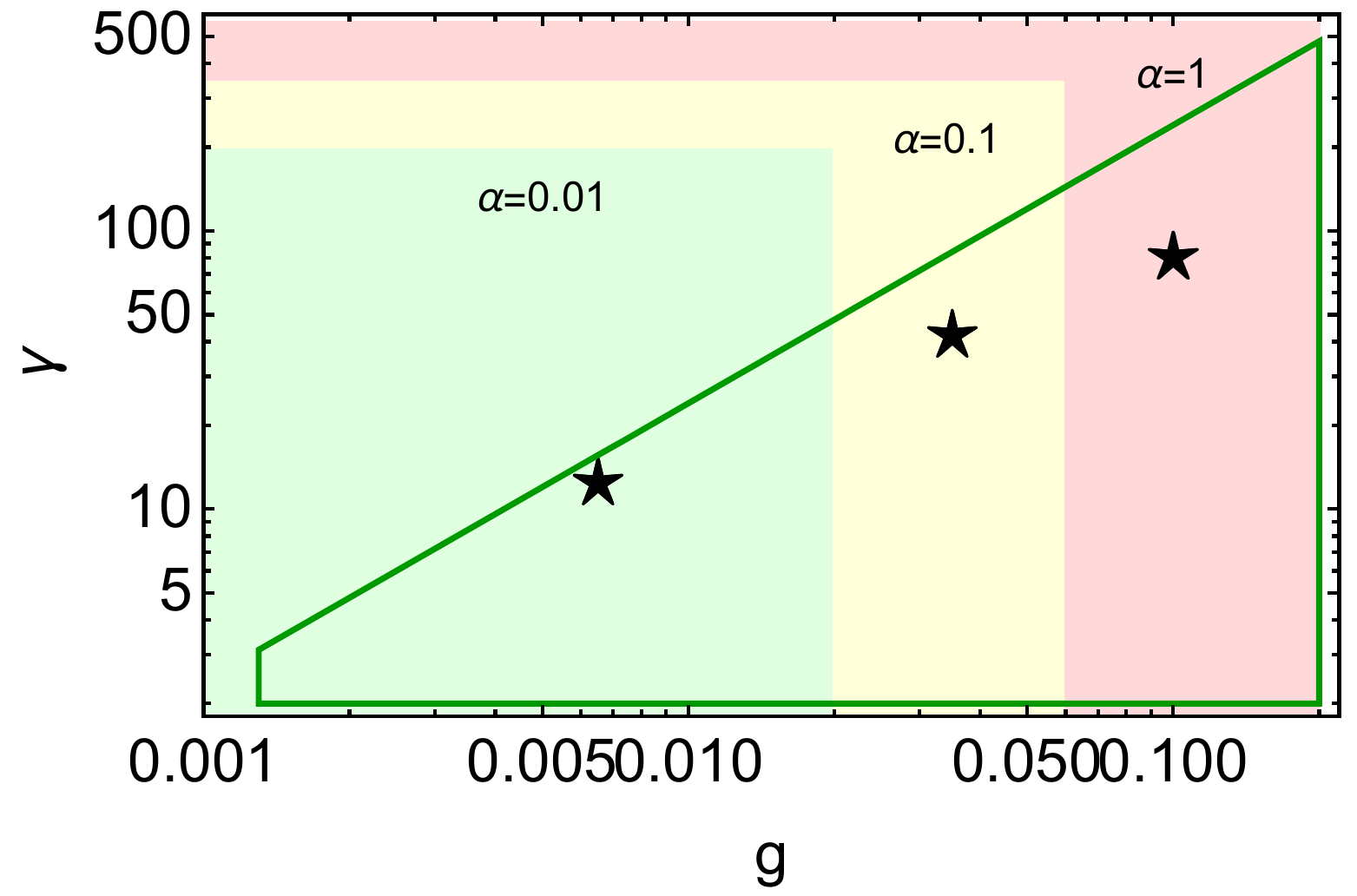}
  \includegraphics[width=0.45\textwidth]{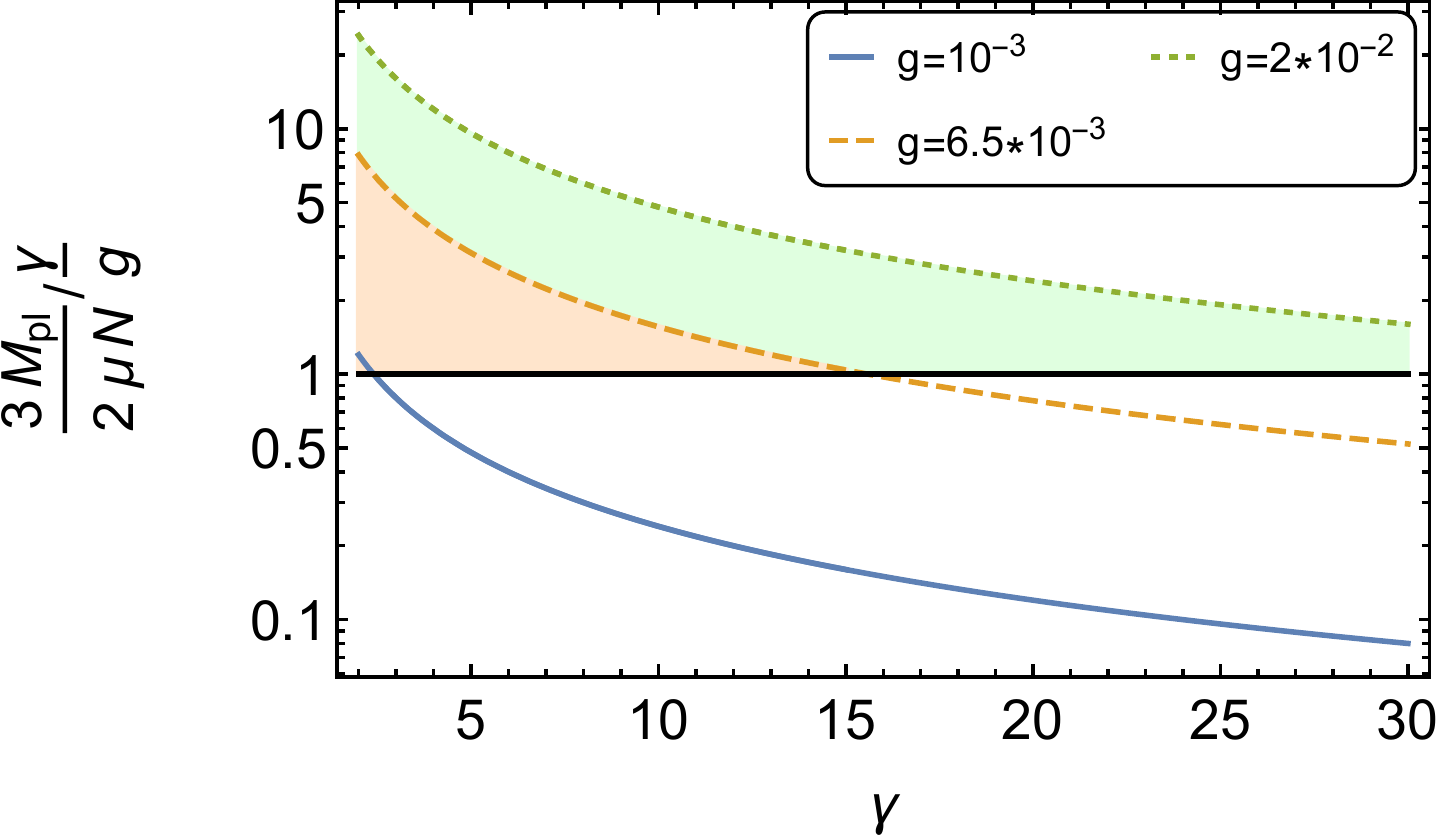}
\caption{\textcolor{black}{
  {\it Left:}  The region plot for $\gamma$ and $g$ which satisfy Eq.~\eqref{epsBcondition} for $N=60$ (solid green line). Shaded colour regions show the theoretically allowed ranges for the gauge coupling g for different $\alpha$, dictated by the requirement for $g_{\rm max}$ (see Eq.~\eqref{loopRequirement}). These allowed regions are: $0.0016\leq g\leq 0.02$ for $\alpha=0.01$ (green),  $0.0016\leq g\leq 0.06$ for $\alpha=0.1$ (yellow) and $0.0016\leq g\leq 0.2$ for $\alpha=1$ (red). Stars show the values that are highlighted in Fig.~\ref{fig:gammaMax} and used in  subsequent simulations.
 }
  {\it Right:}  The ratio $\frac{3 M_{\rm Pl}}{2 \mu N} / \frac{\gamma}{g }$ as a function of $\gamma$ for $N=60$ and $g=10^{-3}, 6.5 \times 10^{-3}, 2\times10^{-2}$ (blue-solid, orange-dashed and green-dotted lines respectively). The black grid lines shows when the ratio equals to 1. 
 }
 \label{fig:RegionPlotGammaG}
\end{figure}

 For given parameters one may numerically evolve the system of Eqs.~\eqref{Hdot} -- \eqref{eomPhi} and find that it is indeed possible to satisfy the conditions of Eqs.~\eqref{rhoEpsilonInequalities1} and
\eqref{rhoEpsilonInequalities2}. Fig.~\eqref{fig:PhiandQ} shows the evolution of the inflaton field $\varphi$ and VEV of the gauge field $Q$ as a function of the number of e-folds $N$. Notice that $Q(N)$ evolves mildly with $N$ and stays almost constant, \textcolor{black}{as implied by $\delta\ll1$}. The evolution of the parameter $\gamma(N)$, which is defined in Eq.~\eqref{delta}, mimics the behaviour of $Q(N)$ which is shown on  Fig. \eqref{fig:PlotsQ}. As we will see in Section \ref{Sec:TensorTilt}, the shape of $\gamma$ determines the tilt of the tensor power spectrum. Since we require that $Q(N)$ also slow-rolls during the slow-roll of $\varphi(N)$, we expect that $\gamma(N)$ to be a decreasing function of time\footnote{The post-inflationary dynamics and the effect of parametric resonance \cite{Iarygina:2018kee, Iarygina:2020dwe, Krajewski:2018moi} is left for future work.}. The evolution of the components of $\epsilon$ and $\rho$ is shown on  Fig.~\eqref{fig:EpsilonRho}. As we have seen in our numerical simulations, for  $\alpha$-attractors the most restrictive condition to host a Gauge-flation sector as a spectator for inflation appears to be the condition $\epsilon_{\varphi}\gg \epsilon_{Q_B}$. It is known  (see e.g. Refs.~\cite{Kallosh:2013yoa, Iarygina:2018kee}), that for $\alpha$-attractors $\epsilon_{\varphi}\simeq\frac{3\alpha}{4N^2}$. Hence, with the definitions of Eqs.~\eqref{epsilon_def} and \eqref{delta}, $\epsilon_{\varphi}\gg \epsilon_{Q_B}$ is satisfied for
\begin{equation}\label{eBcondition}
\frac{3\alpha}{4N^2}\gg \frac{\gamma Q^2}{M_{\rm Pl}^2}.
\end{equation}
By fixing the parameter $\tilde \alpha$, the number of e-folds $N$ and the value of $\gamma>2$, it is easy to find the range of allowed initial values for the gauge field $Q_0$, in order for the non-Abelian sector to stay subdominant. One may rewrite the condition of Eq.~\eqref{eBcondition} using Eq.~\eqref{delta} and $H^2\simeq H_{\varphi}^2\simeq \frac{\alpha \mu^2}{3 }$ in the following form
\begin{equation}
\frac{3 M_{\rm Pl}}{2 \mu N}\gg \frac{\gamma}{g }. \label{epsBcondition}
\end{equation}
Similarly, the condition $\epsilon_{\varphi}\gg \epsilon_{Q_E}$ may be written for $\delta\ll 1$ using Eqs.~\eqref{epsilon_def}, \eqref{delta}
as
\begin{equation}
\frac{3 M_{\rm Pl}}{2 \mu N}\gg \frac{\sqrt{\gamma}}{g }. \label{epsEcondition}
\end{equation}
Indeed, we see that the condition $\epsilon_{\varphi}\gg \epsilon_{Q_B}$ is more restrictive than $\epsilon_\varphi \gg \epsilon_{Q_E}$, which agrees with  our numerical simulations. The left-hand side of Eqs.~\eqref{epsBcondition}, \eqref{epsEcondition} is a fixed number that is set by the number of e-folds of inflation $N$ and the scale $\mu$, that does not depend on the parameters of the potential $\alpha$ and $n$, and is uniquely fixed from the amplitude of the power spectrum of the scalar density perturbations. \textcolor{black}{However, $g_{\rm max}$ depends on the choice of $\alpha$ due to the scaling $H^2 \simeq \frac{\alpha \mu^2}{3 }$ in \eqref{loopRequirement}. To satisfy the above inequality, $g_{\rm max}$ cannot be too large. The range for allowed values for $\gamma$ and $g$ that satisfy Eq.~\eqref{epsBcondition} (for $\kappa$ given by Eq.~\eqref{papametersKappa}) is shown in Fig.~\ref{fig:RegionPlotGammaG}. 
Note, that for  the $\alpha$-attractor model $\gamma$ does not depend on the parameter $\alpha$ due to simultaneous scaling of $H$ and $\epsilon_{\varphi}$ (and hence $\epsilon$) with $\alpha$, but changes significantly with $g$.  }

\section{Tensor sector}\label{Sec:Tilt}

In this Section we  analyze the tensor perturbations generated by the gauge fields. We   identify the parameter space restrictions coming from requiring that the inflaton sector  dominates the energy density and background evolution. These restrictions in turn put constraints on  the gravitational wave production by the gauge sector.

\subsection{Tensor perturbations }
 In this subsection we adopt the notation of Ref.~\cite{Dimastrogiovanni:2016fuu} for tensor perturbations in  the gauge field and the metric.
The tensor sector consists of four independent perturbations that are given by
\begin{gather}
\delta A^1_{\mu}=a(0,t_{+},t_{\times},0), \quad \delta A^2_{\mu}=a(0,t_{\times},-t_{+},0), \quad \delta g_{11}=-\delta g_{22}=a^2h_{+},
\quad \delta g_{12}=a^2h_{\times}.
\end{gather}
The plus and cross polarizations are related to the left-handed and right-handed polarizations as
\begin{gather}\label{LRmodes}
h_{+}=\frac{h_L+h_R}{\sqrt{2}}, \quad h_{\times}=\frac{h_L-h_R}{i\sqrt{2}},\quad t_{+}=\frac{t_L+t_R}{\sqrt{2}},\quad t_{\times}=\frac{t_L-t_R}{i\sqrt{2}}.
\end{gather}
We canonically normalise them by introducing
\begin{equation}
h_{L,R}=\frac{\sqrt{2}}{M_p a}H_{L,R},\quad t_{L,R}=\frac{1}{\sqrt{2}a}T_{L,R},
\end{equation}
 The action for the canonically normalised perturbations reads 
\begin{equation}\label{2ndorderAction}
S_L=\frac{1}{2}\int d\tau d^3k\left[ \Delta'^{\dagger}_L\Delta'_L+\Delta'^{\dagger}_L K_L\Delta_L-\Delta^{\dagger}_LK_L\Delta'_L-
\Delta^{\dagger}_L\Omega^2_L\Delta_L\right], 
\quad \Delta_L=
\begin{pmatrix}
H_{L}\\
T_{L}
\end{pmatrix}.
\end{equation}
where the  expression for the right-handed sector is identical.
Prime $()'$ here denotes a derivative with respect to conformal time $\tau$.  The anti-symmetric matrix $K_{L/R}$ is defined through
\begin{equation}\label{K}
K_{L/R,12}=\frac{1}{M_p}\left(Q'+\frac{a'}{a}Q\right),
\end{equation}
and $\Omega^2_{L/R}$ is symmetric, with components
\begin{gather}\label{Omega}
\Omega^2_{L/R,11}=k^2-2\frac{a'^2}{a^2}+\frac{3g^2a^2Q^4}{M_p^2}-\frac{(aQ)'^2}{M_p^2a^2},\\
\Omega^2_{L/R,12}=\pm k\frac{2gaQ^2}{M_p}+\frac{(aQ)'}{aM_p}\frac{a'}{a}-\frac{2\kappa g^2Q^3}{M_pa^2}\frac{g^2a^4Q^4+a'^2Q^2-a^2Q'^2}{1+\kappa g^2Q^4},\\
\Omega^2_{L/R,22}=k^2 \mp 2kgaQ\left[1+\kappa \frac{g^2a^4Q^4+a'^2Q^2-a^2Q'^2}{a^4(1+\kappa g^2Q^4)}\right]+
\frac{2\kappa g^2Q^2}{a^2}\frac{g^2a^4Q^4+a'^2Q^2-a^2Q'^2}{1+\kappa g^2 Q^4},
\end{gather}
where signs refer to the left-handed or the right-handed polarization respectively, \textcolor{black}{which we denote by `$`L/R$"}. We now use the background relations obtained in  Section~\ref{Sec:BackgroundParameters} to simplify the above matrix elements and expand them in slow-roll in order to identify limitations on the chiral gravitational wave production, coming from the presence of the inflaton field.
It is convenient to rewrite the matrices in terms of $\epsilon, \gamma$ and $\delta$. With substitutions coming from Eqs.~\eqref{delta}, \eqref{kappa} and \eqref{Mpl}
\begin{gather}
Q' \rightarrow -aQH\delta\quad , \quad a' \rightarrow a^2 H \quad , \quad \kappa\rightarrow \frac{1}{H^2\gamma Q^2}\frac{(1-\delta)^2+\gamma}{(1-\delta)^2}\frac{2-\left(\epsilon_Q+\frac{2}{3}\epsilon_{\varphi}+\Upsilon\right)
}{\epsilon_Q},\\
M_p\rightarrow Q\sqrt{\frac{(1-\delta)^2+\gamma}{\epsilon_Q}}\quad , \quad g\rightarrow\sqrt{\gamma}\frac{H}{Q},
\end{gather}
we  find exact expressions for the matrices given by
\begin{gather}
K_{L/R,12}=\frac{aH\sqrt{\epsilon_Q}}{\sqrt{(1-\delta)^2+\gamma}}(1-\delta),\\
\Omega^2_{L/R,11}=k^2-a^2H^2\frac{(1-\delta)^2(2+\epsilon_Q)+\gamma(2-3\epsilon_Q)}{(1-\delta)^2+\gamma},\\
\Omega^2_{L/R,12}=\pm aHk \frac{2\sqrt{\gamma \epsilon_Q}}{\sqrt{(1-\delta)^2+\gamma}}
-a^2H^2\frac{\sqrt{\epsilon_Q}}{\sqrt{(1-\delta)^2+\gamma}}\cdot \\
\cdot\frac{(2\gamma^2+3\gamma(1-\delta))(2-\Upsilon-(\epsilon_Q+2/3\epsilon_{\varphi}))+
(1-\delta)^3(2-\Upsilon-2\epsilon_Q-2/3\epsilon_{\varphi}+2\delta(2-\Upsilon-(\epsilon_Q+2/3\epsilon_{\varphi})))
}
{(2-\Upsilon-2/3\epsilon_{\varphi})(1-\delta)^2+\gamma(2-\Upsilon-(\epsilon_Q+2/3\epsilon_{\varphi}))},\notag \\
\Omega^2_{L/R,22}=k^2\mp2aHk\frac{1}{\sqrt{\gamma}}\cdot \\
\cdot\frac{(2\gamma^2+(1-\delta)^3(1+\delta))(2-\Upsilon-(\epsilon_Q+2/3\epsilon_{\varphi}))+
\gamma(1-\delta)(3(2-\Upsilon)-\delta(2-\Upsilon-2/3\epsilon_{\varphi})-2\epsilon)
}
{(2-\Upsilon-2/3\epsilon_{\varphi})(1-\delta)^2+\gamma(2-\Upsilon-(\epsilon_Q+2/3\epsilon_{\varphi}))}+\notag\\
+2a^2H^2
\frac{(2-\Upsilon-(\epsilon_Q+2/3\epsilon_{\varphi}))(\gamma^2+2(1-\delta)\gamma+(1+\delta)(1-\delta)^3)
}
{(2-\Upsilon-2/3\epsilon_{\varphi})(1-\delta)^2+\gamma(2-\Upsilon-(\epsilon_Q+2/3\epsilon_{\varphi}))}.\notag
\end{gather}

We substitute $\epsilon_Q$ and $\delta$ from Eq.~\eqref{deltasimple}, i.e.
\begin{equation}
\epsilon_Q\rightarrow \epsilon-\epsilon_{\varphi}\quad , \quad
\delta \rightarrow \frac{\epsilon}{3(2-\Upsilon-(\epsilon-\frac{1}{3}\epsilon_{\varphi}))}\left(\epsilon-\frac{2}{3}\frac{\epsilon_{\varphi}}{\epsilon}\eta_{\varphi}-\frac{\epsilon_{\varphi}}{3}+\Upsilon\left(1-\frac{\epsilon_{\varphi}}{\epsilon}\right)\right)
\end{equation}
and expand the matrix elements in slow roll with $\epsilon\ll 1$ and $\epsilon_{\varphi}\ll 1$.
The lowest order in slow-roll quantities is $\sqrt{\epsilon}$, where we obtain
\begin{gather}\label{matrixC}
K_{L/R,12}\simeq aH\frac{\sqrt{\epsilon}}{\sqrt{1+\gamma}}\,
C_1(\epsilon_{\varphi}),\\
\Omega^2_{L/R,11}\simeq k^2 -2a^2H^2,\\
\Omega^2_{L/R,12}\simeq \left( \pm 2kaH\frac{\sqrt{\gamma\epsilon}}{\sqrt{1+\gamma}}-a^2H^2\frac{1+2\gamma}{\sqrt{1+\gamma}}\sqrt{\epsilon}\right)C_1(\epsilon_{\varphi}),\\
\Omega^2_{L/R,22}\simeq  k^2\mp 2kaH\frac{1}{\sqrt{\gamma}}\left[ 1+2\gamma+C_2(\epsilon_{\varphi}) \right]+
2a^2H^2\left[ 1+\gamma+C_2(\epsilon_{\varphi}) 
\right],
\end{gather}
where we introduced the ``correction'' coefficients 
\begin{gather}
C_1(\epsilon_{\varphi})=\sqrt{1-\frac{\epsilon_{\varphi}}{\epsilon}},\\
C_2(\epsilon_{\varphi})=-\frac{\epsilon}{2-\Upsilon}\left(1-\frac{\epsilon_{\varphi}}{\epsilon}\right)+\cal{O}\left(\epsilon\right).
\end{gather}
Notice that when the inflaton field dominates the energy budget,  Eq.~\eqref{HSquare} leads to $\Upsilon \sim 2$. Hence the second correction becomes $\frac{\epsilon}{2-\Upsilon}\left(1-\frac{\epsilon_{\varphi}}{\epsilon}\right)\sim\sqrt{\epsilon}$ and is taken into account for consistency. For the case $\epsilon_{\varphi}=0$ and $\Upsilon=0$, matrix elements reduce to the case of pure Gauge-flation  and agree with the results obtained in Ref.~\cite{Dimastrogiovanni:2016fuu}. The absolute value of the corrections $C_1(\epsilon_{\varphi})$ and $C_2(\epsilon_{\varphi})$ depend on the fraction of energy stored in the inflaton field, i.e. $\epsilon_{\varphi}/\epsilon$. 
There is an interesting ``tug of war'' between two different effects here. 
\begin{itemize}
\item Since $1-\epsilon_\varphi/\epsilon = \epsilon_Q/\epsilon$, GW production by the gauge sector requires $\epsilon_\varphi/\epsilon$ to deviate somewhat from unity.   
\item The requirement that the gauge-sector does not affect the dynamics of inflation and the generation of density fluctuations is encoded in $\epsilon_{\varphi}\gg \epsilon_Q$ or $\epsilon_{Q}/\epsilon \ll 1$.
\end{itemize}
Both requirements, the dominance of the inflaton sector and significant GW production by the gauge sector, can be simultaneously satisfied, but limit the available  parameter space. \textcolor{black}{In particular, the ratio $\epsilon_\varphi/\epsilon$ determines the allowed range of values for the parameter $\gamma$, as seen from Eq.~\eqref{trueGammaMax} and illustrated in Fig.~\ref{fig:gammaMax} and Fig.~\ref{fig:RegionPlotGammaG}. In turn, $\gamma$ controls the amplification of chiral gravitational waves with respect to the  vacuum gravitational wave value. At the same time Eq.~\eqref{rhoEpsilonInequalities2} must be satisfied
for the gauge sector to be a spectator.
% We illustrate that it is possible to achieve this in Fig. \ref{fig:PlotsQ}. 
 \textcolor{black}{Fig. \ref{fig:PlotsQ} shows that these  conditions can be simultaneously satisfied.}  
 \\
For $\epsilon_\varphi/\epsilon \simeq 0.99$ the requirement $\epsilon_{\varphi}\gg \epsilon_Q$ is satisfied automatically. However this choice will result in small allowed values for $\gamma$ and hence in a small amplification of the gravitational wave background, making it observationally uninteresting. 
It is therefore important for the ratio $\epsilon_\varphi/\epsilon$   to deviate from unity for the system to exhibit a significant sourcing of gravitational waves.
\\
At the same time, the ratio $\epsilon_{\varphi}\gg \epsilon_Q$ cannot be   too small, in order to ensure that the gauge sector is a spectator. 
\textcolor{black}{We numerically checked that for $\epsilon_\varphi/\epsilon\gtrsim 0.6$ the energy density contribution of the gauge sector is kept at the percent level or below.}
}

The equation of motion for tensor perturbations follows from Eq.~\eqref{2ndorderAction} and may be written in the form
\begin{equation}
\Delta''_L+2K_L \Delta_L ' +(K'_L +\Omega^2_L)\Delta_L=0,
\end{equation}
and similarly for the right-handed sector.
To  leading order in $\sqrt{\epsilon}$ and neglecting interactions with the gravitational wave sector, the equation of motion for the gauge field perturbation reads
\begin{equation}
\partial^2_{\tau} T_{L} +\Omega^2_{L,22}T_{L}=0.
\end{equation}
Substituting the matrix $\Omega^2_{L,22}$ explicitly with $\tau=-\frac{1}{aH}$ we get
\begin{equation}\label{TL}
\partial^2_{\tau}T_{L} + \left( k^2 -  \frac{2k}{-\tau}\frac{1+2\gamma+C_2(\epsilon_{\varphi}) }{\sqrt{\gamma}}+
 \frac{2\left( 1+\gamma+C_2(\epsilon_{\varphi}) \right)}{\tau^2} \right)T_{L}=0.
\end{equation}
Now, we may define $z=2ik\tau$ and
\begin{gather}
\tilde{\nu}=\frac{2\left(1+2\gamma+C_2(\epsilon_{\varphi}) \right)}{\sqrt{\gamma}}=-2i\tilde{\alpha}, \\
\tilde{\mu}=2\left(1+\gamma+C_2(\epsilon_{\varphi}) \right)=\frac{1}{4}-\tilde{\beta}^2,
\end{gather}
in order to rewrite Eq.~\eqref{TL} in the form of the Whittaker equation
\begin{equation}
\partial^2_{z} T_{L} + \left( -\frac{1}{4}+\frac{\tilde{\alpha}}{z}+\frac{\frac{1}{4}-\tilde{\beta}^2}{z^2} \right)T_{L}=0.
\end{equation}
The solution is expressed in terms of  Whittaker functions as
\begin{equation}\label{T0}
T_{L,0}(k,\tau)=A_k M_{\tilde{\alpha},\tilde{\beta}}(2ik\tau)+B_k W_{\tilde{\alpha},\tilde{\beta}}(2ik\tau),
\end{equation}
with $ M_{\tilde{\alpha},\tilde{\beta}}(2ik\tau)$ and $W_{\tilde{\alpha},\tilde{\beta}}(2ik\tau)$ being the Whittaker  M and W functions. Here the subscript $0$ indicates that we neglected interactions with the gravitational wave sector. 
In the asymptotic past $x\equiv-k\tau \rightarrow \infty$, the solution approaches the Bunch-Davies vacuum, i.e.
\begin{equation}
T_{L,0}(k,\tau) \rightarrow \frac{1}{\sqrt{2k}}e^{ix}.
\end{equation}
Asymptotic expansions for the Whittaker functions in this limit are also well-known, hence the constants $A_k$ and $B_k$ in \eqref{T0} are given by \cite{Adshead:2017hnc}
\begin{gather}\label{AB}
A_k=\frac{1}{\sqrt{2k}} \frac{\Gamma\left(-\tilde{\alpha}+\tilde{\beta}+\frac{1}{2}\right)}{(2i)^{-\tilde{\alpha}}\Gamma\left(2\tilde{\beta} +1\right)},\\
B_k=\frac{1}{\sqrt{2k}}\frac{\Gamma\left(-\tilde{\alpha}+\tilde{\beta}+\frac{1}{2}\right)}{\Gamma\left(\tilde{\alpha}+\tilde{\beta}+\frac{1}{2}\right)}
2^{\tilde{\alpha}}i^{\tilde{\beta}+1}(-i)^{\tilde{\alpha}-\tilde{\beta}}.
\end{gather}

\begin{figure}
\centering
 \includegraphics[width=0.45\textwidth]{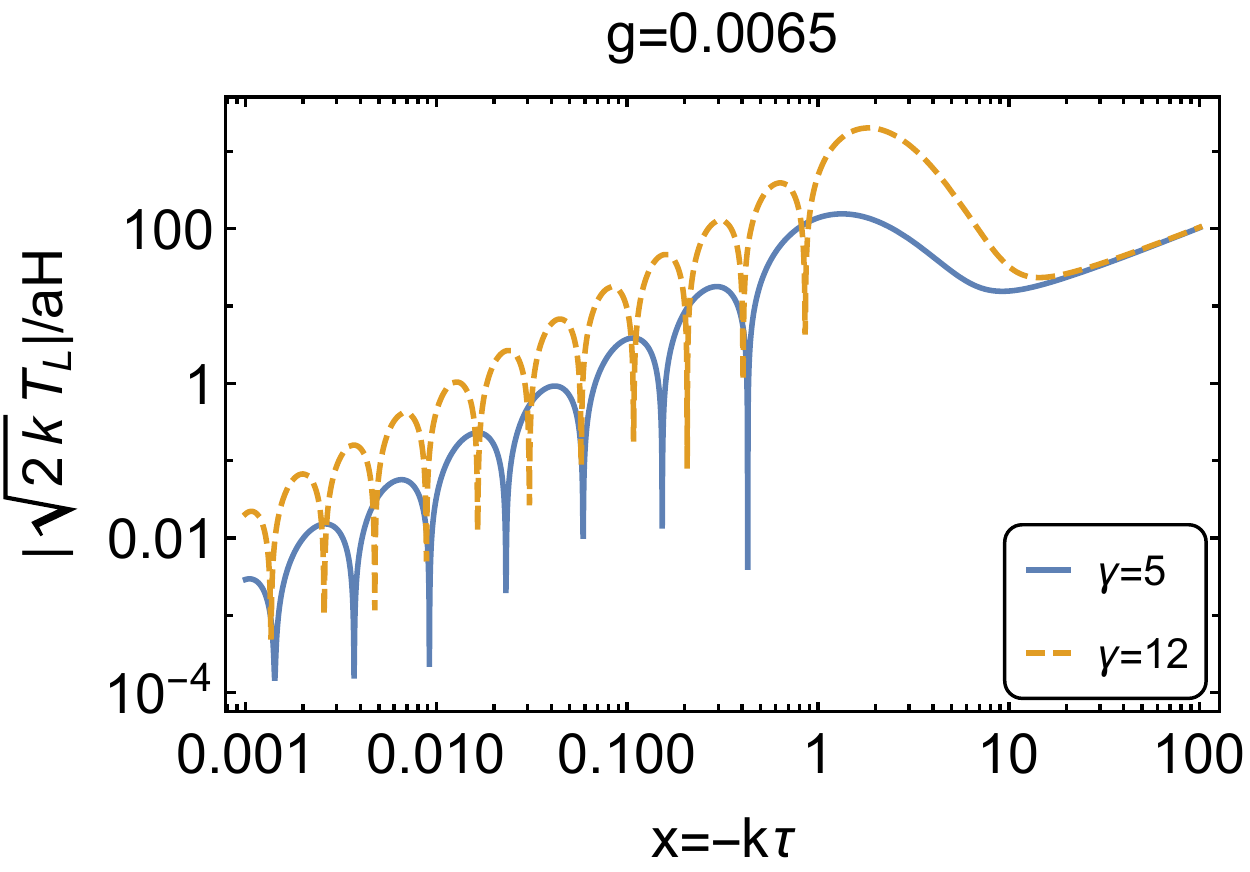}
\includegraphics[width=0.435\textwidth]{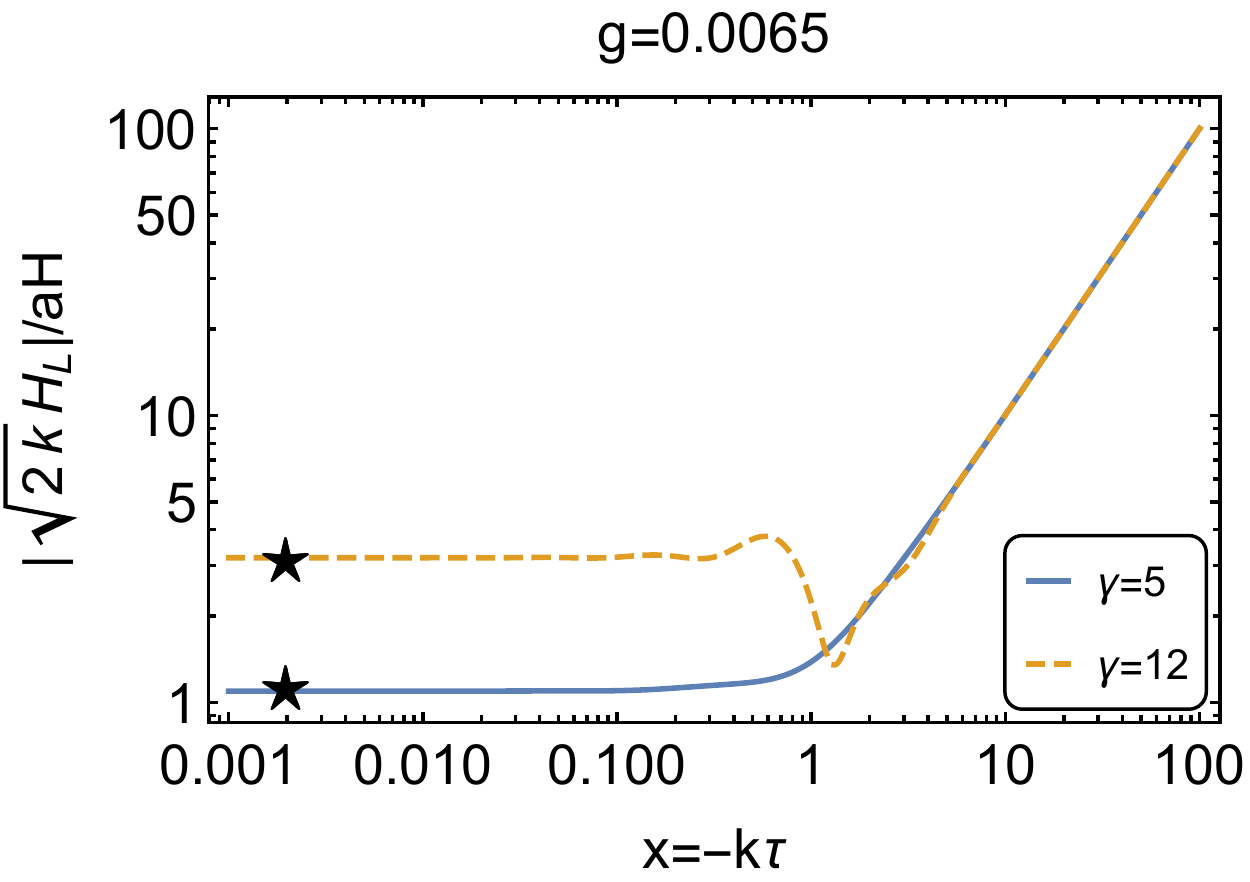}\\
 \includegraphics[width=0.45\textwidth]{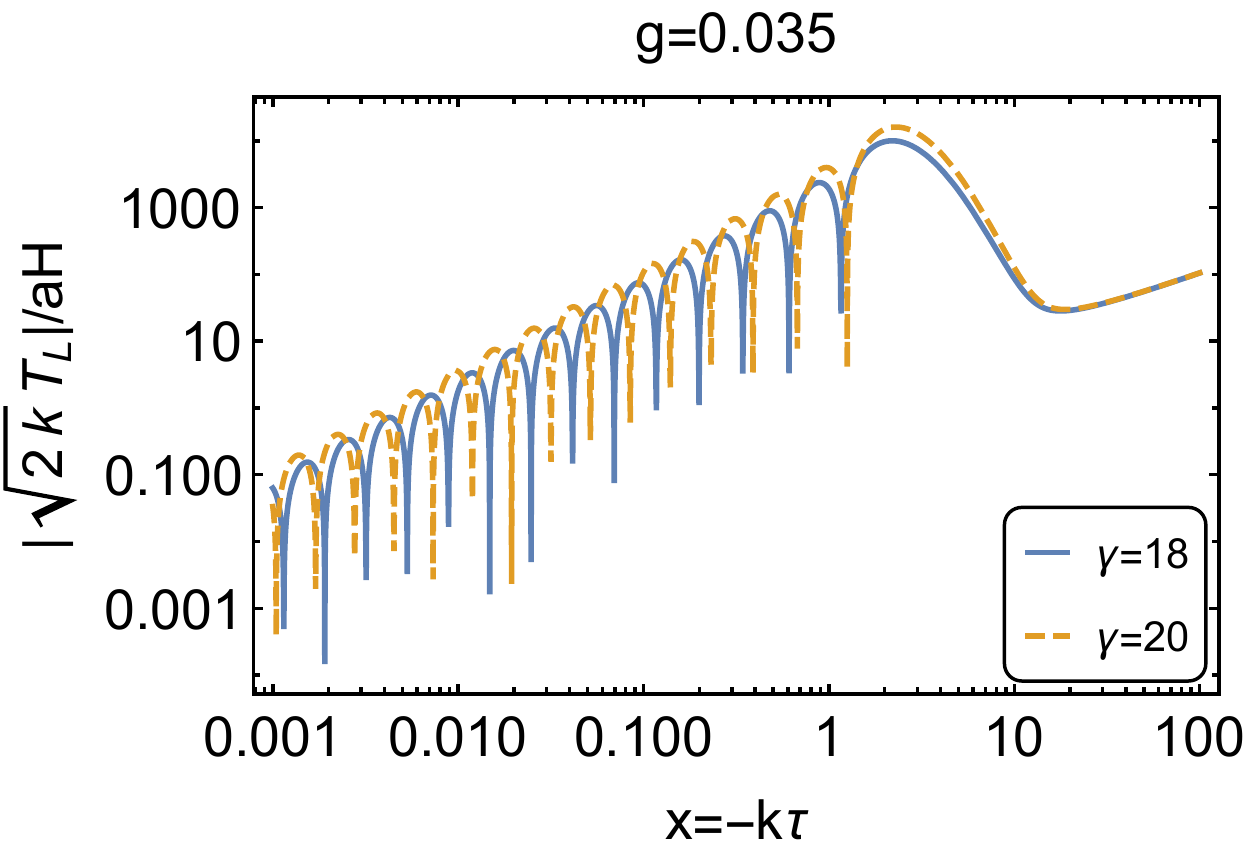}
\includegraphics[width=0.435\textwidth]{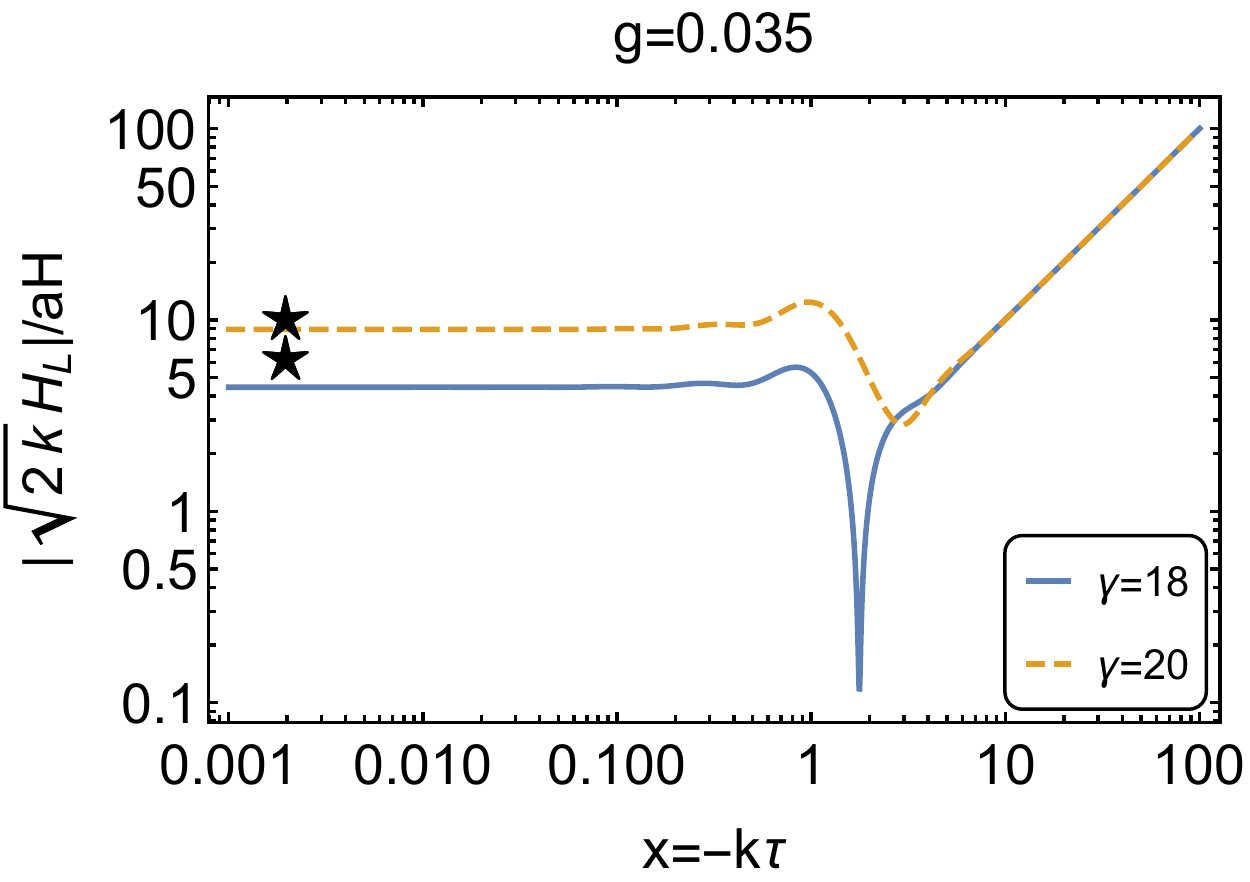}
\caption{ \textcolor{black}{
 {\it Left panels:} Tachyonic growth of the left-polarized gauge field mode-function $T_L$ around the time of horizon crossing $x=1$ for $H/M_{\rm pl}=1.9*10^{-6}$ (that corresponds to $\alpha$-attractor model with $\alpha=0.1$) for values of $g$ and $\gamma$. On \textit{the top panel} the parameters are: $g=0.0065$ and $\gamma=5,12$ (blue solid and orange dashed lines respectively) and on\textit{ the bottom panel}: $g=0.035$ and $\gamma=18, 20$ (blue solid and orange dashed lines respectively). Note that  we only use   values of $\gamma$ that are compatible with the observational constraints for the chosen values of $g$ and $\alpha$, as discussed in Sec.~\ref{Sec:TensorTilt} (see Fig.~\ref{fig:r}). For the bottom panel these values are smaller than the maximally theoretically allowed quantities shown as stars in Fig.~\ref{fig:gammaMax}.
 {\it Right panels:} Enhancement of the left-polarized GW mode-function $H_L$, sourced by the gauge field mode-function $T_L$ for the same parameters and color coding as on the corresponding left panels. Stars represent the approximate late-time solutions given by Eq.~\eqref{HLlate}. 
\textcolor{black}{ The plots are normalized such that the late-time value of the right-polarized vacuum GW mode-function equals unity.}
 }
 }
 \label{fig:TLHL}
\end{figure}

Next, we find that metric tensor modes to leading order in $\sqrt{\epsilon}$ satisfy the following equation of motion in the $x$-variable
\begin{equation}\label{gammaL}
\partial^2_{x} H_L + \left( 1-\frac{2}{x^2}\right)H_L=
\frac{\sqrt{\epsilon}\,C_1(\epsilon_{\varphi})}{\sqrt{1+\gamma}}\left( \frac{2}{x}\partial_x T_L +\left(\frac{2\gamma}{x^2}-\frac{2\sqrt{\gamma}}{x}\right)T_L\right).
\end{equation}
Using the Born approximation, one may find the solution of Eq.~\eqref{gammaL} in series of $\sqrt{\epsilon}$
\begin{equation}
H_L=H_{L,0} +H_{L,s},
\end{equation}
where $H_{L,0}$ is the homogeneous solution of the free equation of motion, and  $H_{L,\text{s}}$ is inhomogeneous part that is sourced by the gauge filed perturbation $T_L$. The homogeneous solution matches the Bunch-Davies vacuum at asymptotic past and is given by 
\begin{equation}
H_{L,0}=\frac{1}{\sqrt{2k}}\left(1+\frac{i}{x}\right)e^{ix}.
\end{equation}
The sourced piece of the solution may be written as
\begin{equation}\label{HS}
H_{L,s}=\frac{\sqrt{\epsilon}\,C_1(\epsilon_{\varphi})}{\sqrt{1+\gamma}}
\int^x dx'
\left( \frac{2}{x'}\partial_{x'}  +\left(\frac{2\gamma}{x'^2}-\frac{2\sqrt{\gamma}}{x'}\right)\right)G(x,x')T_{L,0} (x'),
\end{equation}
where $G(x,x')$ is the Green's function. We can follow the same steps as in Refs.~\cite{Adshead:2013nka, Adshead:2016omu, Adshead:2017hnc} and find that the late-time solution for the left-handed gravitational wave is given by
\begin{equation}\label{HLlate}
H_L=\frac{Hx}{M_{\text{Pl}}\sqrt{k^3}}u_1(x)+2\sqrt{2}\frac{H}{M_{\text{Pl}} k}B_k\frac{2\sqrt{\epsilon} \, C_1(\epsilon_{\varphi})}{\sqrt{1+\gamma}}\left(I_1+\sqrt{\gamma}I_2-\gamma I_3\right), 
\end{equation}
which contains a free and a sourced part of the solution. Here we have defined 
\begin{equation}
u_1(x)\equiv \left(1+\frac{i}{x}\right)e^{ix}.
\end{equation}
The terms $I_1, I_2, I_3$ are coming from the integrals in Eq.~\eqref{HS} and expressed as 
\begin{gather}
I_1=\frac{\left( \tilde{\mu}^2-2i\tilde{\mu}\tilde{\nu}+2\tilde{\mu}-2\tilde{\nu}^2\right)\sec(\pi \tilde{\beta})\sinh(-i\pi\tilde{\alpha})\Gamma(\tilde{\alpha})}{2\tilde{\mu}(\tilde{\mu}+2)}\\
-\frac{\pi^2 \left(\tilde{\mu} ^2 +2i\tilde{\mu}\tilde{\nu}+2\tilde{\mu}-2\tilde{\nu} ^2\right)\sec(\pi \tilde{\beta})\text{csch}(-i\pi\tilde{\alpha})}{
2\tilde{\mu}(\tilde{\mu}+2)\Gamma(\tilde{\alpha}+1)\Gamma(-\tilde{\alpha}-\tilde{\beta}+\frac{1}{2})\Gamma(-\tilde{\alpha}+\tilde{\beta}+\frac{1}{2})},\notag \\
I_2=\frac{\pi\sec(\pi \tilde{\beta})\Gamma(-\tilde{\alpha})}
{2\Gamma(-\tilde{\alpha}-\tilde{\beta}+\frac{1}{2})\Gamma(-\tilde{\alpha}+\tilde{\beta}+\frac{1}{2})}-
\frac{\pi\sec(\pi \tilde{\beta})\Gamma(1-\tilde{\alpha})}
{\tilde{\mu}\Gamma(-\tilde{\alpha}-\tilde{\beta}+\frac{1}{2})\Gamma(-\tilde{\alpha}+\tilde{\beta}+\frac{1}{2})}\\
+\frac{\pi\tilde{\mu}\sec(\pi \tilde{\beta})-i\pi\tilde{\nu}\sec(\pi \tilde{\beta})}
{2\tilde{\mu}\Gamma(1-\tilde{\alpha})},\notag \\
I_3=\frac{\pi^2\left(\tilde{\mu}+i\tilde{\nu}\right)\sec(\pi \tilde{\beta})\text{csch}(-i\pi\tilde{\alpha})}
{\tilde{\mu}(\tilde{\mu}+2)\Gamma(\tilde{\alpha})\Gamma(-\tilde{\alpha}-\tilde{\beta}+\frac{1}{2})\Gamma(-\tilde{\alpha}+\tilde{\beta}+\frac{1}{2})}+
\frac{\pi\left(\tilde{\nu}+i\tilde{\mu}\right)\sec(\pi \tilde{\beta})}{\tilde{\mu}(\tilde{\mu}+2)\Gamma(-\tilde{\alpha})}.
\end{gather}
The homogeneous solution for the gauge field perturbation $T_{L,0}$ is an excellent approximation, since it breaks down for $x\lesssim 0.1$, which does not influence gravitational wave modes which are sourced around horizon crossing $x\simeq 1$. %Indeed, we see that the late-time solution of Eq.~\eqref{HLlate} is in a remarkable agreement with full numerical simulations, as seen on  Fig.~\ref{fig:TLHL}. 
Indeed,  Fig.~\ref{fig:TLHL} shows remarkable agreement between the late-time solution of Eq.~\eqref{HLlate} and the corresponding  full numerical simulations.

The right-handed  gravitational waves do not get enhanced and are given by the usual vacuum value
\begin{equation}
H_R(x)=\frac{Hx}{M_{\text{Pl}}\sqrt{k^3}}u_1(x).
\end{equation}
Finally, the power spectra for left-handed modes can be written as
\begin{equation}\label{PL}
P_{L}^2(k)=\frac{H^2}{2\pi^2 M_{\text{Pl}}^2}+\frac{16kH^2}{\pi^2M_{\text{Pl}}^2}
\frac{\epsilon \,C_1^2(\epsilon_{\varphi}) }{1+\gamma}|B_k|^2\left|I_1+\sqrt{\gamma}I_2-\gamma I_3\right|^2.
\end{equation}
The power spectra for right-handed modes is
\begin{equation}
P_{R}^2(k)=\frac{H^2}{2\pi^2 M_{\text{Pl}}^2}.
\end{equation}
The total tensor power spectrum is given by
\begin{equation}\label{PTtotal}
P_T(k)=2P_{L}^2(k)+2P_{R}^2(k),
\end{equation}
which in the limit $\epsilon\rightarrow\epsilon_{\varphi}$, i.e. $C_1(\epsilon_{\varphi})\rightarrow 0$, reduces to the ``standard'' inflationary result
% single scalar field result 
\begin{equation}\label{PT0}
P_{T,\varphi}(k)=\frac{2H^2}{\pi^2 M_{\text{Pl}}^2},
\end{equation}

Finally, we can define the chirality parameter as
\begin{equation}
\Delta \chi=\frac{P_{L}^2-P_{R}^2}{P_{L}^2+P_{R}^2} \, ,
\end{equation}
\begin{figure}[h!]
\centering
 \includegraphics[width=0.45\textwidth]{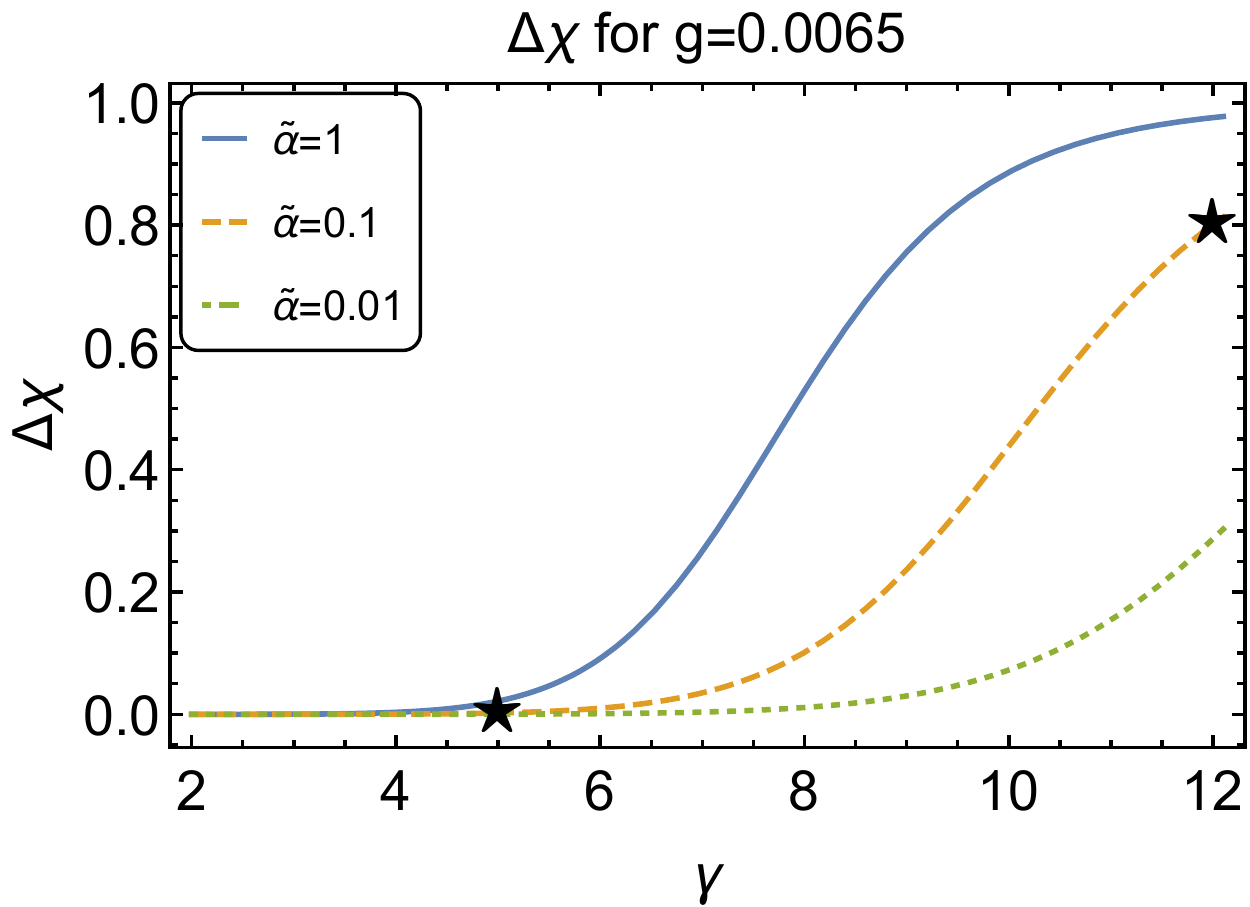}
  \includegraphics[width=0.45\textwidth]{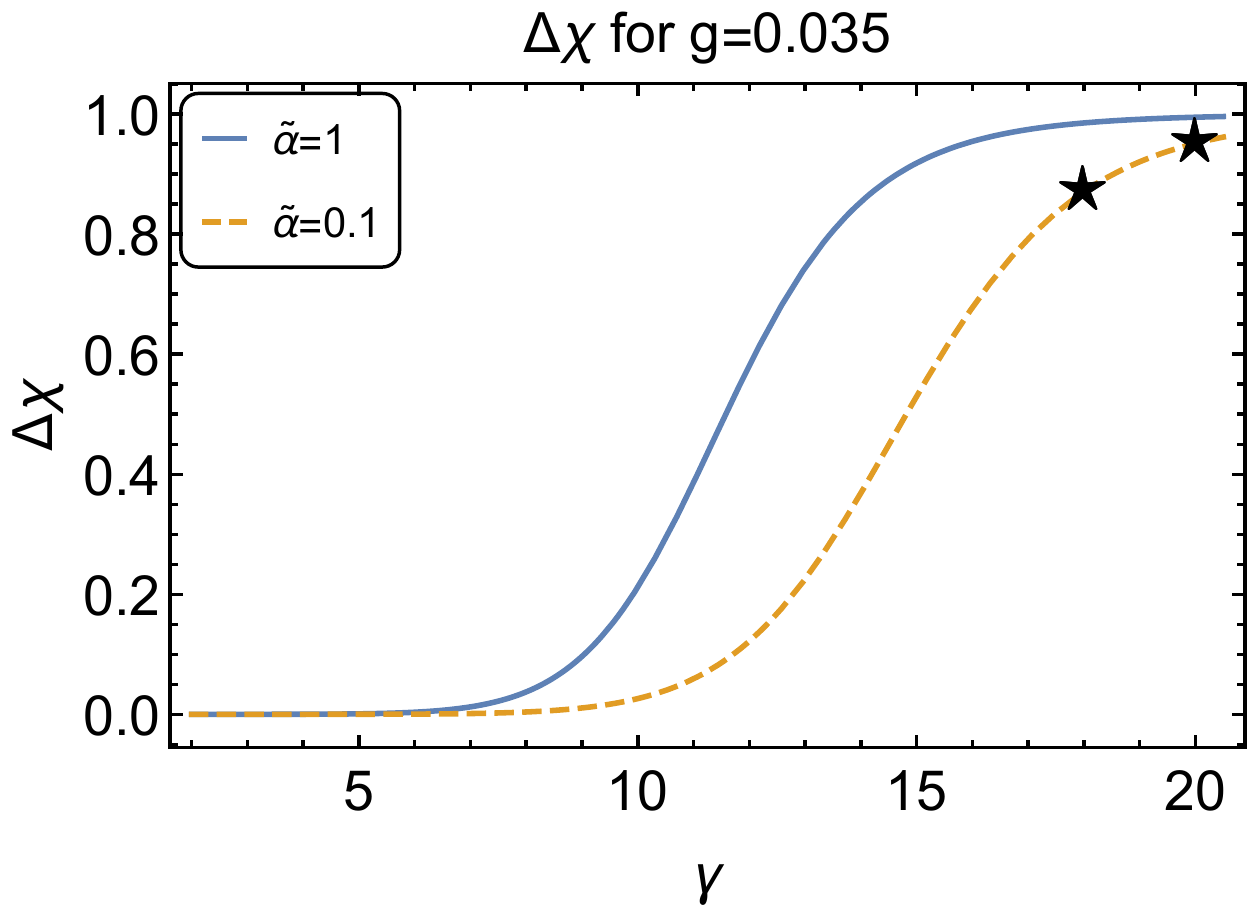}
\caption{\textcolor{black}{
The chirality parameter $\Delta \chi$ as a function of $\gamma$ for $\alpha=1, 0.1, 0.01$ (blue solid, orange dashed and green dot-dashed lines respectively) for $g=0.0065, 0.035$ (\textit{left} and \textit{right} panels respectively). Stars represent the values of $\gamma$ used in Fig. \eqref{fig:TLHL} for the corresponding $\alpha$ and gauge couplings $g$: $\gamma=5, 12$ on the left panel, $\gamma=18, 20$ on the right panel.
 }
 }
 \label{fig:chirality}
\end{figure}
\textcolor{black}{which is shown on Fig.~\ref{fig:chirality}. We see that sufficient enhancement of one of the polarizations occurs for $\gamma\gtrsim 8$ for $g=0.0065$ (left panel) and for $\gamma\gtrsim 10$ for $g=0.035$ (right panel).  }

\subsection{Tensor tilt}\label{Sec:TensorTilt}

In this subsection we  discuss the  shape of the tensor power spectrum generated by  spectator Gauge-flation, characterized by the tensor tilt $n_T$. In Ref.~\cite{Fujita:2018ndp}, it was shown that  spectator Chromo-natural inflation, depending on the choice of the axion potential, supports both flat, red and blue tilted tensor spectra. Thus, our primary interest is to investigate if spectator Gauge-flation may generate all three possible values of the tensor tilt in realistic physical set-ups. The tensor tilt for Eq.~\eqref{PT0} is given by $n_{T}=-2\epsilon_*$, where $\epsilon$ is evaluated at $t=t_*$ that defines time of horizon crossing for a mode with the wave number $k_*=a(t_*)H$. Below we  focus on the tilt for the sourced part only.

The power spectra of sourced gravitational waves from Eqs.~\eqref{PL} and \eqref{PTtotal} are given by
\begin{equation}\label{PTs}
P_{T,s}(k)=\frac{32 kH^2}{\pi^2M_{\text{Pl}}^2}
\frac{\epsilon \,C_1^2(\epsilon_{\varphi}) }{1+\gamma}|B_k|^2\left|I_1+\sqrt{\gamma}I_2-\gamma I_3\right|^2.
\end{equation}
We are going to proceed as follows: first we  rewrite Eq.~\eqref{PTs} in terms of $\gamma(t)$, restoring its time-dependence, and then re-express $P_{T,s}^2(k)$ in terms of $\gamma(k)$. This allows us to calculate the tensor tilt.

The time evolution of the vacuum expectation value of the gauge field $Q(t)$ may be written as
\begin{equation}
Q(t)=Q(t_*)+\dot{Q}(t_*)(t-t_*),
\end{equation}
with $t_*$ being the time of horizon crossing.
From here it follows that
\begin{equation}
\frac{Q(t)}{Q(t_*)}=1-\delta_* H(t-t_*),
\end{equation}
where $\delta_*=-\frac{\dot{Q}(t_*)}{HQ(t_*)}$. This gives the time dependence of the parameter $\gamma(t)$, i.e.
\begin{gather}
\gamma(t)=\gamma_*\left( \frac{Q(t)}{Q(t_*)}\right)^2=\gamma_*\left(1-\delta_* H(t-t_*)\right)^2 \simeq \gamma_*\left( 1+2\frac{H(t-t_*)}{\Delta N}\right),
\end{gather}
with $\gamma_*=\frac{g^2Q^2(t_*)}{H^2}$, $\Delta N=-1/\delta_*$.
Using $H(t-t_*)=\ln(k/k_*)$ we can write $\gamma(k)$ as
\begin{equation}
\gamma(k)\simeq\gamma_*\left( 1+2\frac{\ln(k/k_*)}{\Delta N}\right) \simeq \gamma_* e^{ \left(\frac{2\ln(k/k_*)}{\Delta N}\right) } .
\end{equation}

To start with, using $|\Gamma(\frac{1}{2}+i b)|^2=\frac{\pi}{\cosh(\pi b)}$ one can rewrite $|B_k|^2$ defined in Eq.~\eqref{AB} in terms of $\gamma(t)$ as
\begin{equation}
|B_k|^2=\frac{1}{2k} e^{3\pi \left(\frac{1+2\gamma}{\sqrt{\gamma}}\right)}e^{-\pi \sqrt{7+8\gamma} }\,
\frac{1+e^{-\pi\left( \sqrt{7+8\gamma}+\frac{2(1+2\gamma)}{\sqrt{\gamma} } \right)  }}{1+e^{-\pi\left( \sqrt{7+8\gamma}-\frac{2(1+2\gamma)}{\sqrt{\gamma} } \right)  }}.
\end{equation}
Next, from Eq.~\eqref{epsilon1} for $\delta\ll 1$ one can find
\begin{equation}
\epsilon \,C_1^2(\epsilon_{\varphi}) =\epsilon_Q\simeq \frac{H^2}{g^2 M_{\text{Pl}}^2} \gamma (1+\gamma).
\end{equation}
The term $\gamma I_3$ generates the main contribution in Eq.~\eqref{PTs}, hence we will neglect smaller contributions coming from $I_1, \sqrt{\gamma}I_2$. In terms of $\gamma$ we find
\begin{equation}\label{I3}
\gamma^2 \left| I_3\right|^2=\gamma^2 \frac{\pi \left(1+2\gamma)(1+\gamma(1+\gamma)(5+\gamma)\right)}{\gamma^{3/2}(1+\gamma)^2(2+\gamma)^2}
e^{-\pi \left( \sqrt{7+8\gamma}-\frac{1+2\gamma}{\sqrt{\gamma}}\right)}.
\end{equation}
Putting everything together, the sourced tensor power spectrum becomes
\begin{gather}
P_{T,s}(k)\simeq\frac{16 H^4}{\pi^2 g^2 M_{\text{Pl}}^4}\gamma \, e^{3\pi \left(\frac{1+2\gamma}{\sqrt{\gamma}}\right)}e^{-\pi \sqrt{7+8\gamma} }\,
\frac{1+e^{-\pi\left( \sqrt{7+8\gamma}+\frac{2(1+2\gamma)}{\sqrt{\gamma} } \right)  }}{1+e^{-\pi\left( \sqrt{7+8\gamma}-\frac{2(1+2\gamma)}{\sqrt{\gamma} } \right)  }}\left|\gamma I_3\right|^2,
\end{gather}
where $\left|\gamma I_3\right|^2$ is given by Eq.~\eqref{I3}.
Next, one may expand
\begin{gather}
\sqrt{7+8\gamma}=\sqrt{7+8\gamma_*}+\frac{8\gamma_*}{\sqrt{7+8\gamma_*}\Delta N}\ln\left(\frac{k}{k_*}\right)-\frac{32\gamma_*^2}{(7+8\gamma_*)^{3/2}(\Delta N)^2}\ln^2\left(\frac{k}{k_*}\right),\\
\frac{1+2\gamma}{\sqrt{\gamma}}=\frac{1+2\gamma_*}{\sqrt{\gamma_*}}+\frac{2\gamma_*-1}{\sqrt{\gamma_*}\Delta N}\ln\left(\frac{k}{k_*}\right)-\frac{\gamma_*-3/2}{\sqrt{\gamma_*}(\Delta N)^2}\ln^2\left(\frac{k}{k_*}\right).
\end{gather}
We do not present the final expression for $P_{T,s}(k)$ in terms of $\gamma_*, k, k_*$ since it is rather cumbersome, but may be easily derived from the above expressions. We  focus instead on the tensor tilt. As usual, the tensor power spectra may be written in the form
\begin{equation}
P_{T,s}(k)=A_T(\gamma_*)\left(\frac{k}{k_*}\right)^{n_{T,s}},
\end{equation}
 with the tensor tilt  given by
\begin{align}
n_{T,s}  =
\frac{d \ln P_{T,s}(k)}{d \ln k}=&
-\delta_* \,[3+4\pi \frac{2\gamma_*-1}{\sqrt{\gamma_*}}-16 \pi \frac{\gamma_*}{\sqrt{7+8\gamma_*}}
\nonumber
\\
&-\frac{2\gamma_*(-8-39\gamma_*-57\gamma_*^2-23\gamma_*^3+\gamma_*^4)}{(1+\gamma_*)(2+\gamma_*)(1+2\gamma_*)(1+5\gamma_*+6\gamma_*^2+\gamma_*^3)}
\nonumber
\\
&-\pi\frac{\frac{8\gamma_*}{\sqrt{7+8\gamma_*}}+2\frac{2\gamma_*-1}{\sqrt{\gamma_*}} }
{1+e^{\pi\left(\sqrt{7+8\gamma_*}+2 \frac{1+2\gamma_*}{\sqrt{\gamma_*}} \right)}}+
\pi\frac{\frac{8\gamma_*}{\sqrt{7+8\gamma_*}}-2\frac{2\gamma_*-1}{\sqrt{\gamma_*}} }
{1+e^{\pi\left(\sqrt{7+8\gamma_*}-2 \frac{1+2\gamma_*}{\sqrt{\gamma_*}} \right)}}
 ],
\end{align}
where we neglected ${\cal O} \left(\delta^2_*\right)$ corrections  and ignored the time-dependence of $H$. For $\alpha$-attractors, as well as other plateau models, this is a very good approximation.
The complicated expression above may be fitted via a simple  formula
\begin{equation}\label{nT}
n_{T,s}\simeq  -\delta_*  \left(3+1.225 \pi \frac{2\gamma_*-1}{\sqrt{\gamma_*}}-3.612\pi \frac{\gamma_*}{\sqrt{7+8\gamma_*}}\right)
\simeq  -\delta_*  \left(2.85+3.68 \sqrt{ \gamma_*}\right).
\end{equation}
Hence, we may conclude that if $Q(t)$ is a decreasing function of time, then $\delta(t)$ defined in Eq.~\eqref{delta} leads to  $\delta_*$ being positive. Therefore, as follows from Eq.~\eqref{nT}, a red-tilted power spectrum is generated. If, on the contrary,  $Q(t)$ increases in time, $\delta(t)$ is  negative, which sources a blue-tilted spectrum. We can sum up the relation of $Q(t)$ to $n_T$ in the following table
\begin{equation}
\label{eq:QandnT}
\begin{split}
Q(t)\searrow \quad &\Rightarrow \quad \dot{Q}(t)<0 \quad \Rightarrow \quad \delta>0 \quad \Rightarrow \quad  n_T<0 \quad \text{red tilt},\\
Q(t)\nearrow \quad &\Rightarrow \quad \dot{Q}(t)>0 \quad \Rightarrow \quad \delta<0 \quad \Rightarrow \quad  n_T>0 \quad \text{blue tilt}.
\end{split}
\end{equation}

All the results shown in Section \ref{Sec:Viability} contain  $Q(t)$ as a decreasing function of time, leading to red-tilted tensor spectra. 
Finally,  \textcolor{black}{by considering only the inflaton-generated scalar fluctuations $P_\zeta$,} Eq.~\eqref{PTtotal} leads to the tensor-to-scalar ratio $r$
\begin{equation}
r=\frac{P_T}{P_{\zeta}}.
\end{equation}

\begin{figure}
\centering
 \includegraphics[width=0.45\textwidth]{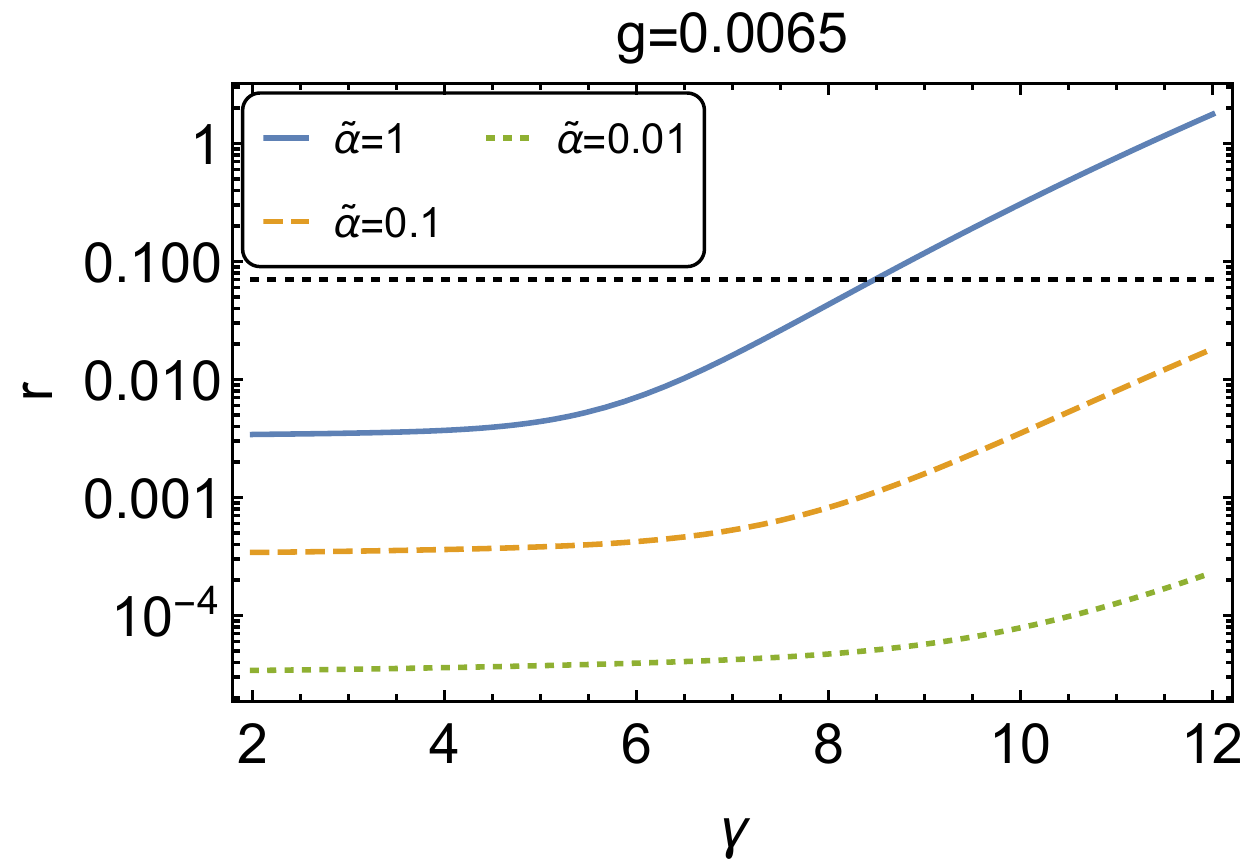}
  \includegraphics[width=0.45\textwidth]{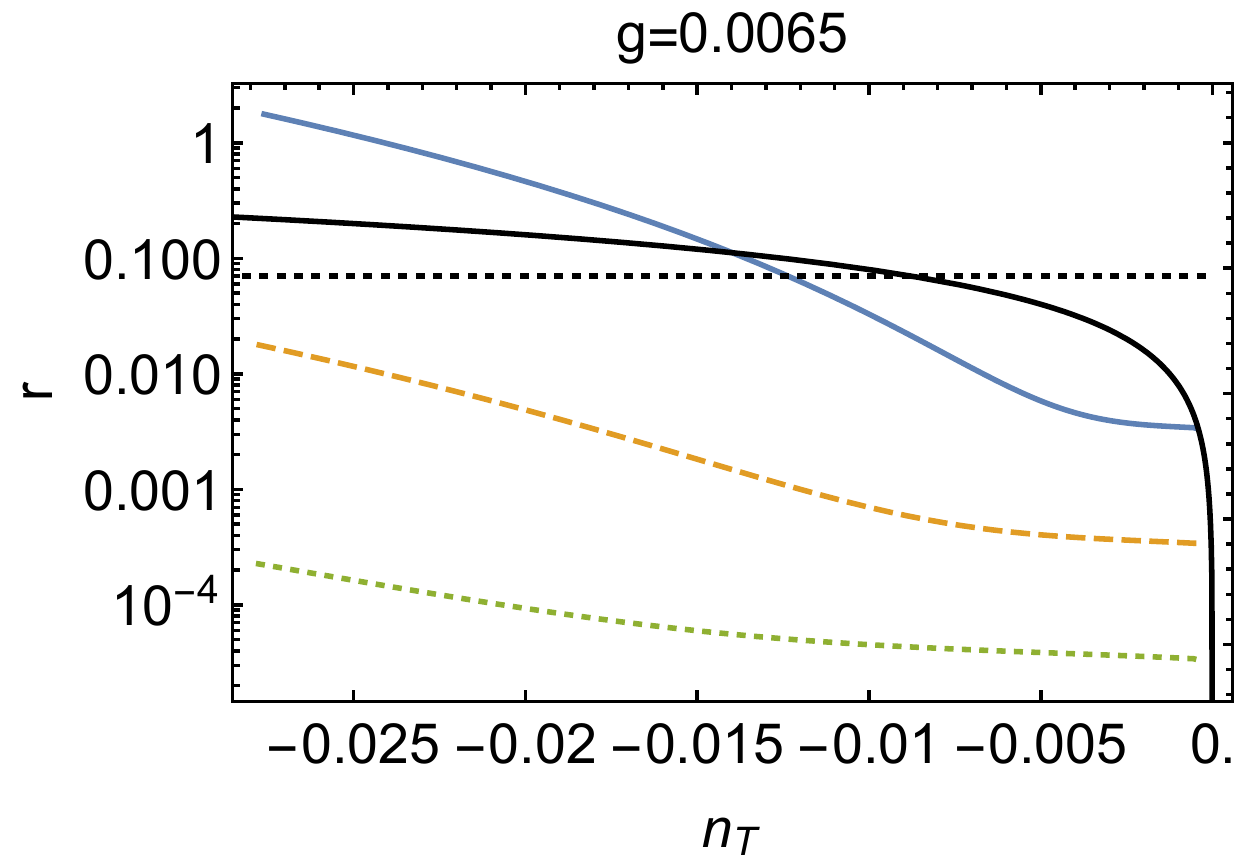}\\
   \includegraphics[width=0.45\textwidth]{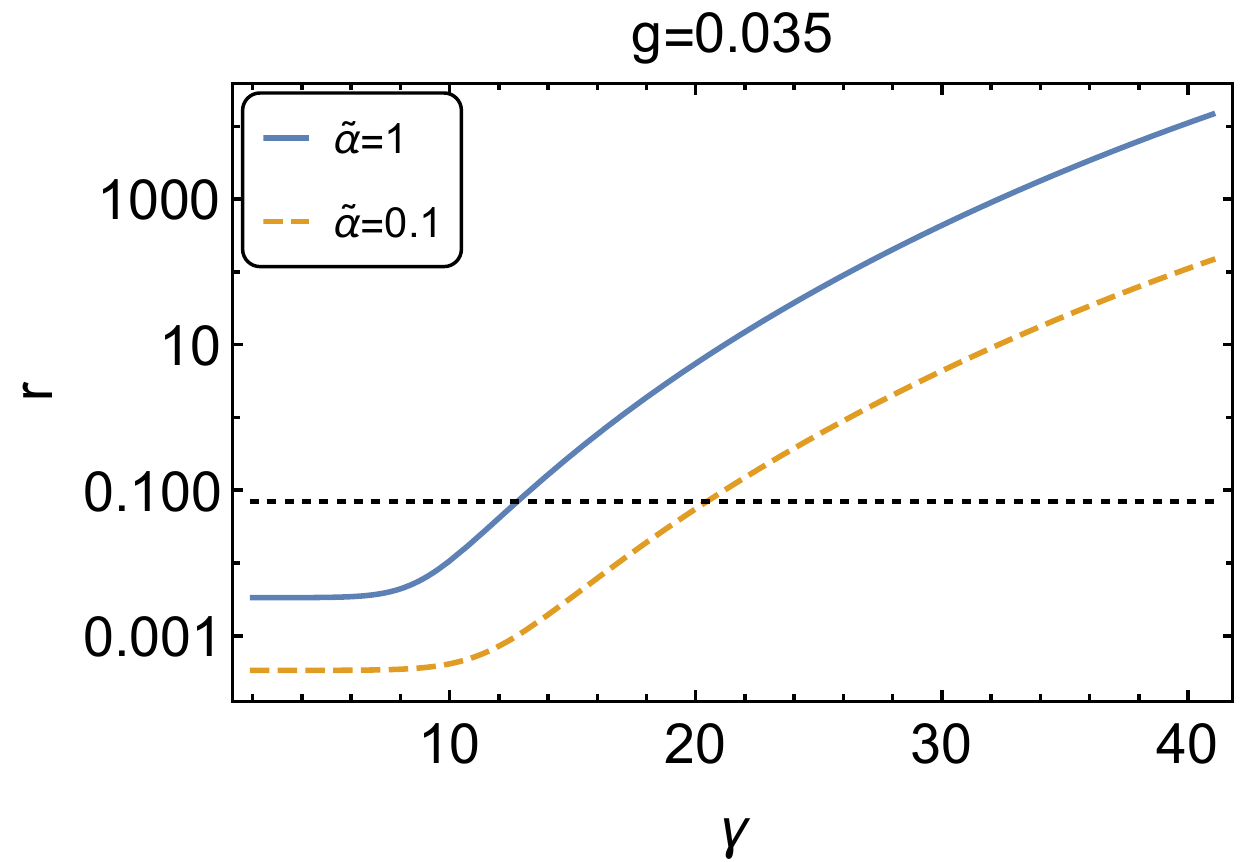}
  \includegraphics[width=0.45\textwidth]{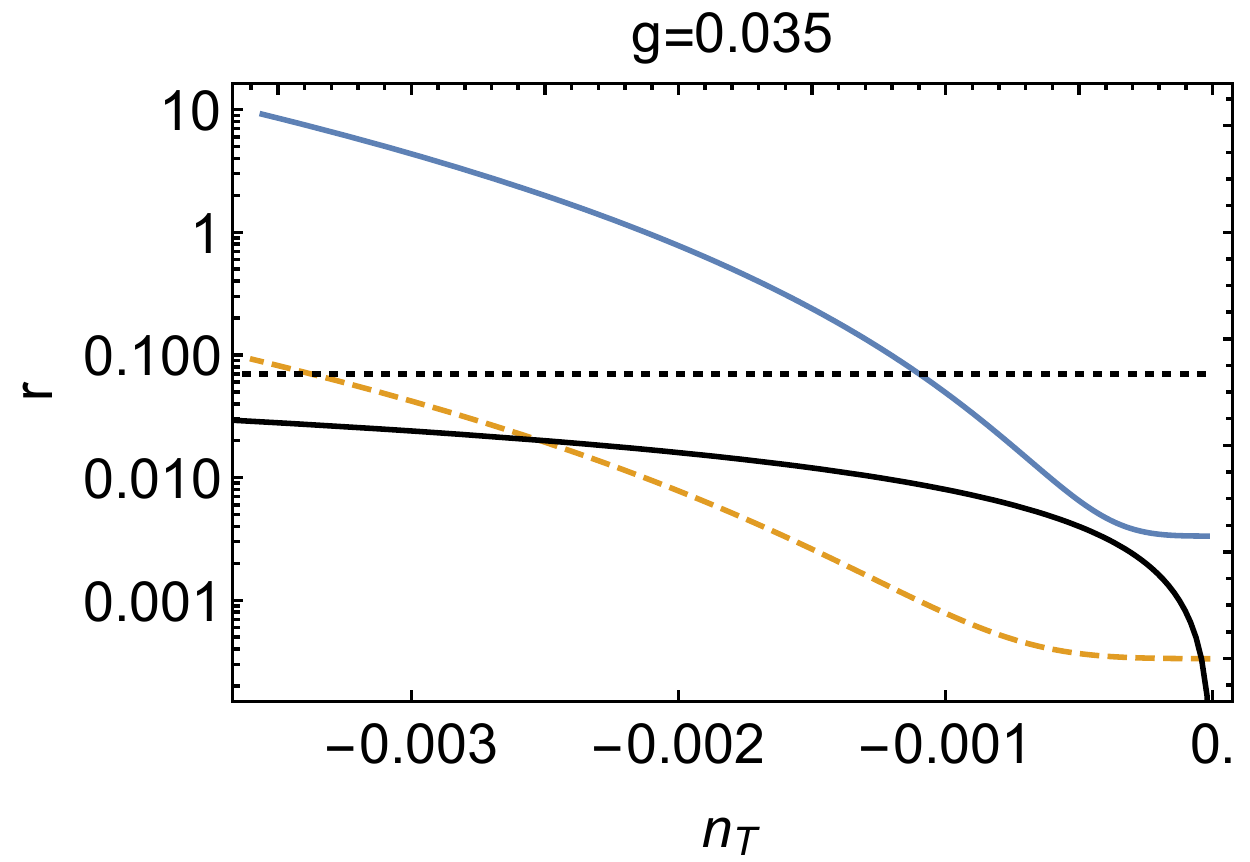}\\
   \includegraphics[width=0.45\textwidth]{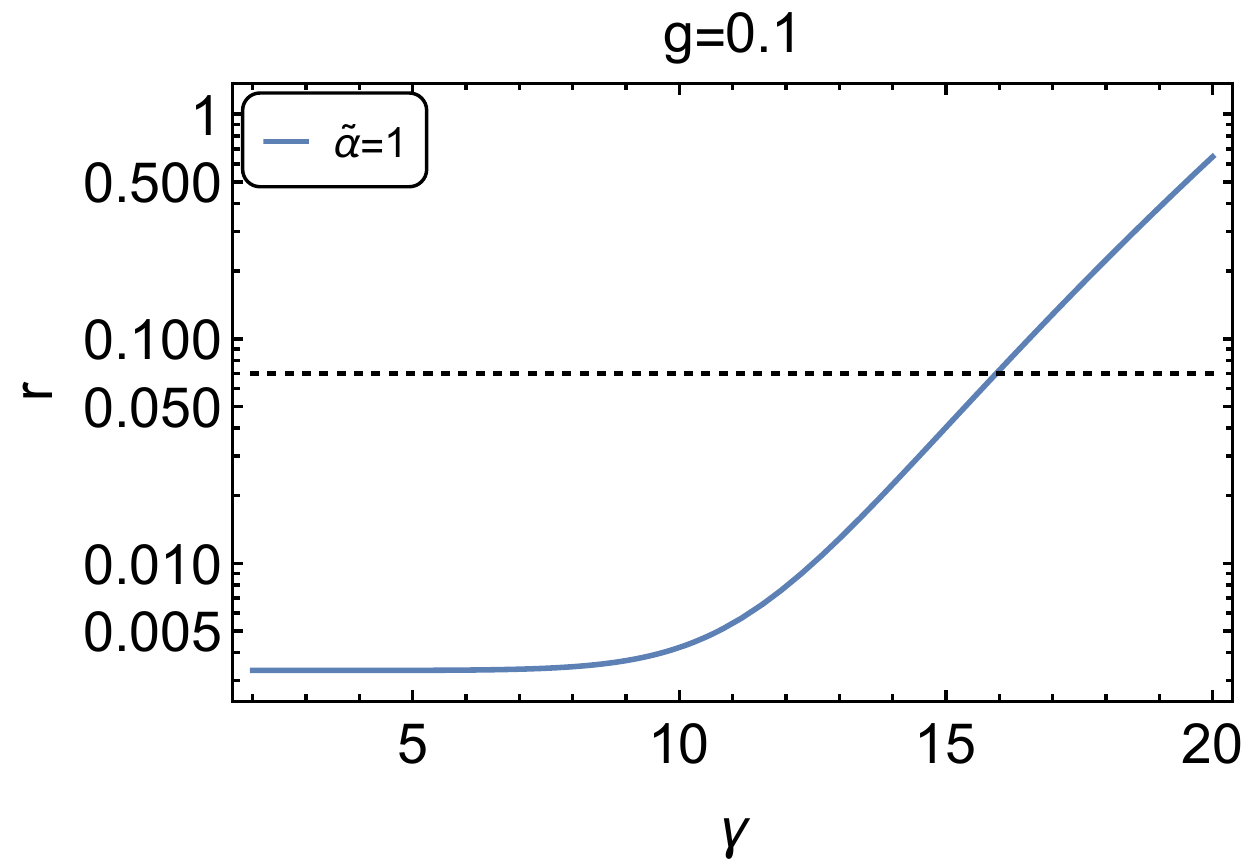}
  \includegraphics[width=0.45\textwidth]{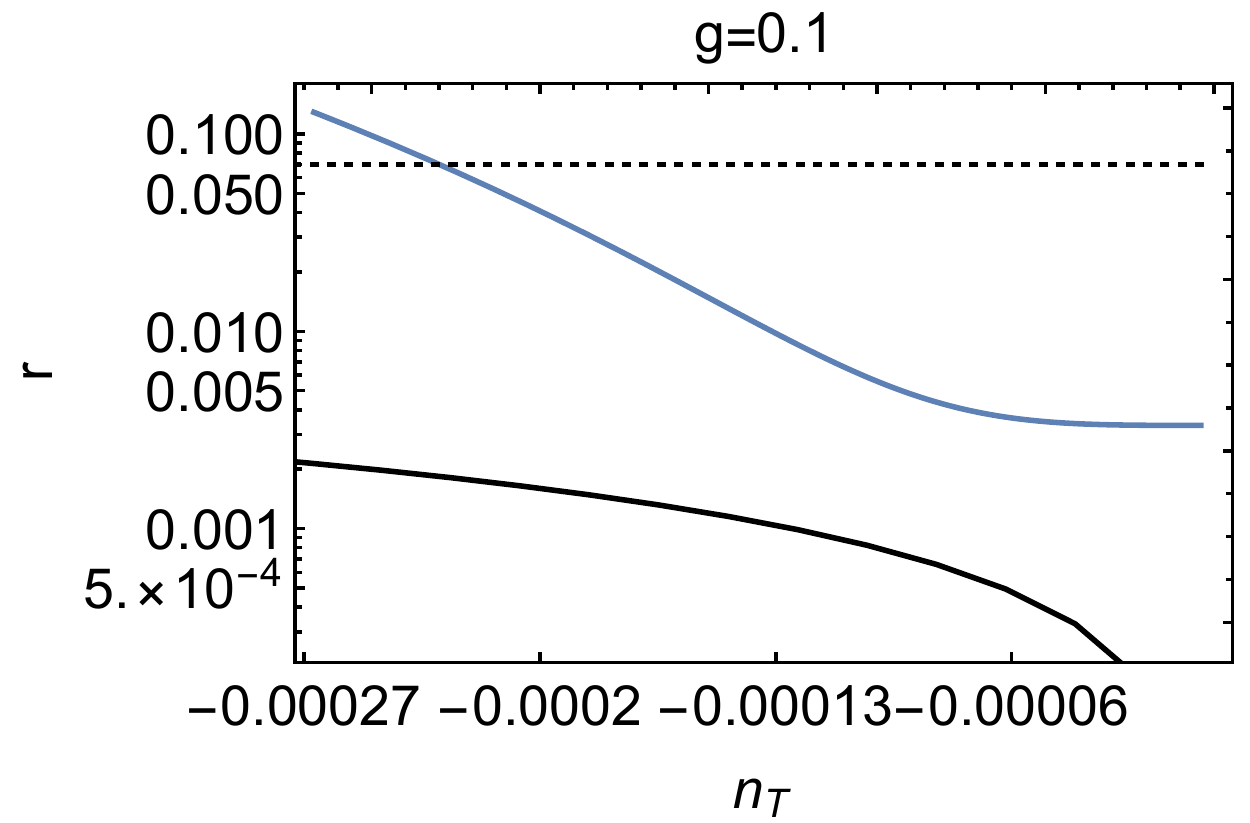}
 \caption{\textcolor{black}{
{\it Left:} The tensor-to-scalar ratio $r$ as a function of $\gamma$ for $\alpha= 1, 0.1, 0.01$ (blue solid, orange dashed and green dotted lines respectively) and $g=0.0065$ (\textit{top panel}),  $g=0.035$ (\textit{middle panel}), $g=0.1$ (\textit{bottom panel}). In each panel we  only show the allowed values of $\alpha$ for the corresponding $g$, according to Fig.~\ref{fig:RegionPlotGammaG} .
For upper bounds on $\gamma$ we use values marked with stars in Fig.\ref{fig:gammaMax} with the corresponding ratios of $\epsilon_{\varphi}/\epsilon$. Parameters $H$ and  $\epsilon_{\varphi}$ in Fig.\ref{fig:gammaMax} correspond to the case $\alpha= 1$. For the $\alpha$-attractor model $\gamma$ does not depend on $\alpha$ due to the scaling of $H$ and   $\epsilon_{\varphi}$ with $\alpha$; however it depends strongly on $g$. The horizontal black-dotted curve shows the observational upper limit on $r$. 
{\it Right:} The tensor-to-scalar ratio $r$ versus the tensor tilt $n_T$ for the same parameters $g, \alpha$ and color coding as on the  left. The ranges for $\gamma$ are: $\gamma\leq 12$ (top) $\gamma\leq 21$  (middle) and $\gamma\leq 17$ (bottom).
These values are chosen in order to meet the observational limits dictated by the corresponding left panels. We can see the range of $n_T$ and the clear departure from the single field consistency relation $n_T=-r/8$ (solid black curve).
 } 
 }
 \label{fig:r}
\end{figure}

\textcolor{black}{The left panel of Fig.~\ref{fig:r} shows the enhancement of the tensor-to-scalar ratio $r$ and its dependence on $\gamma$ for the $\alpha$-attractor potential of Eq.~\eqref{Tpotential} with $n=3/2$, $\alpha= 1, 0.1, 0.01$ and gauge couplings $g=0.0065, 0.035, 0.1$. Note that we show only values of $\alpha$ which are allowed for a given gauge coupling $g$, in order to satisfy the $\kappa$-term dominance over the loop corrections required by Eq.~\eqref{loopRequirement}, see also the left panel of Fig. \ref{fig:RegionPlotGammaG}.
We see that for small $\gamma$, we recover the single field $\alpha$-attractor result $r=16\epsilon$ with $\epsilon\to\epsilon_{\varphi}\simeq{3\alpha}/{(4N^2)}$.
Further increasing $r$ requires decreasing the gauge coupling $g$. However this is severely restricted by Eqs.~\eqref{gmin} and \eqref{epsBcondition}, meaning that we cannot increase $r$ significantly above what is shown on Fig.~\ref{fig:r}.
The right panel of Fig.~\ref{fig:r} shows the correlation of Eq.~\eqref{nT} and  $r$ using Eq.~\eqref{deltasimple}. We see that $0>n_T\gtrsim -0.03$ and larger $r$ correlates with more red-tilted spectra.}

Before we proceed to a brief overview of related models and comparison with our results on spectator gauge-flation, it is worth discussing the conditions for a red-tilted spectrum. It was shown in Ref.~\cite{Maleknejad:2012fw} that the original gauge-flation model can lead (at the background level) to both decreasing and growing functions of $Q(t)$, depending on the initial conditions. Trajectories starting close to the slow-roll attractor lead to a decreasing $Q(t)$. Trajectories that start far from the slow roll attractor in Ref.~\cite{Maleknejad:2012fw} were shown to undergo a brief period of $\epsilon>1$, followed by a slow-roll inflationary phase with $Q(t)$ increasing in time. The latter behavior required different ranges of $\kappa$ and $g$. 

We were able to recover this general trend in our spectator model, at the cost of altering the parameter space of the model. In particular, to produce a growing $Q(t)$ and a correspondingly blue-tilted GW spectrum, we need to increase the value of $\kappa$. This leads to an increase in $\rho_{\kappa}$, which is bounded from above by the requirement $\rho_{\kappa} \ll \rho_\varphi$. Furthermore, $\gamma$ is reduced for these trajectories, suppressing GW production by the spectator sector. In order to increase $GW$ production, we need to increase $g$, which cannot be done arbitrarily. Such a realization of spectator gauge-flation is given in Appendix~\ref{app:blueGW}. Our numerical tests have shown the existence of such solutions, but at the same time an increased level of parameter fine-tuning is needed to achieve them, at least in the context of an $\alpha$-attractor inflationary sector. We thus consider the red-tilted GW spectra as a ``generic'' prediction of spectator gauge-flation, keeping in mind the ability of these models to evade this prediction for proper choices of parameters and initial conditions. We leave an exhaustive parameter search for a variety of inflationary sectors for future work.

\subsection{Comparison with related models}

The model presented here is part of a larger family of inflationary models, where the existence of a non-abelian sector leads to the generation of chiral GWs. We can distinguish between the original models, where the $SU(2)$ or axion$-SU(2)$ sectors are responsible for inflation and the generation of both scalar and tensor modes, and the spectator models, where the inflaton sector is decoupled from the non-abelian spectator sector. Gauge-flation and Chromo-natural inflation, along with their Higgsed variants, fit in the first category, while spectator Chromo-natural inflation and spectator Gauge-flation make up the second category.

While the original Chromo-natural inflation and Gauge-flation models are ruled out by observations, their Higgsed counterparts provide predictions compatible with  CMB observations for some part of parameter space.
Interesting features arise from the correlation of the resulting tensor to scalar ratio $r$ and tensor spectral tilt $n_T$. For Higgsed gauge-flation, Ref.~\cite{Adshead:2017hnc} showed a negative correlation between the two quantities. For $n\lesssim 0.01$ a blue-tilted spectrum is preferred. Hence Higgsed gauge-flation and spectator gauge-flation (with an $\alpha$-attractor inflationary sector) tend to provide opposite predictions for the sign of $n_T$.

Higgsed chromo-natural inflation has an interesting space of predictions for $n_T$ and $r$. For smaller values of $r<0.01$, the correlation between $r$ and $n_T$ is also mostly negative. However the possible range of values for $n_T$ is much larger, ranging between $-0.2<n_T<0.05$ for the parameter scan presented in Ref.~\cite{Adshead:2016omu}. This means that the possible range of values for Higgs Chromo-Natural inflation is significantly larger than that of our realization of spectator Gauge-flation (see Fig.~\ref{fig:r}) for red and blue tilted spectra alike. 

Next, we wish to compare the present model to spectator Chromo-natural inflation, in which an axion$-SU(2)$ spectator sector is added to an otherwise dominant inflaton. The existence of an axion potential $V(\chi)$ leads for significant diversity in the form of the tensor power spectrum. Ref.~\cite{Fujita:2018ndp} showed the emergence of a blue or red-tilted spectrum for monomial potential $V(\chi)\propto \left |\chi\right|^p$, where the tensor tilt scales as $n_T \propto (p-1)$. A linear potential leads to an exactly scale-invariant tensor spectrum. In principle, small deviations from $p=1$ can lead to an arbitrarily small tensor tilt. However, this requires a rather fine-tuned axion potential. If instead we look at $p=1/2$ and $p=3/2$ we can see that $n_T\simeq 0.04$ and $n_T\simeq-0.07$ respectively. Concave potentials lead to blue-tilted spectra, which are generically not produced in our model of spectator Gauge-flation. On the other hand, convex potentials lead to red-tilted spectra, but for $p> 3/2$ the spectral tilt will be $n_T\lesssim {\cal O}(0.1)$, which is outside the predictions shown in Fig.~\ref{fig:r}.

Finally, Ref.~\cite{Maleknejad:2016qjz} studied an axion-inflaton field coupled to an $SU(2)$ gauge field, where the VEV of the latter is not large enough to affect the background inflationary dynamics.  
Despite being subdominant, the presence of the gauge field can lead to the enhancement of GW's and the corresponding violation of the Lyth bound. The resulting tensor tilt  $n_T$ exhibits oscillations in time and asymptotes to zero at late times.

\section{Summary and discussion}\label{Sec:Summary}

In this work we have explored the phenomenology of  Gauge-flation as a spectator sector during inflation. We have uncovered significant parameter restrictions, arising both from the physics of the gauge sector as well as from the requirements that the gauge sector be subdominant to the inflationary sector. Most importantly, these requirements lead to significant constraints on the parameter $\gamma$, which controls the amount of GW enhancement. 

By identifying the inflationary sector with a realization of the well-known T-model of $\alpha$-attractors, we showed that a spectator gauge-flation sector can increase the tensor-to-scalar ratio by two orders of magnitude. The resulting tensor spectral index $n_T$ is controlled by the evolution of the gauge field vacuum expectation value $Q(t)$, being red if $Q$ is a decreasing function of time during inflation and blue otherwise. The majority of our numerical simulations resulted in red-tilted GW spectra with \textcolor{black}{$-0.03\lesssim n_T <0 $. }

Our work presents an interesting generalization of gauge-flation, while opening up exciting possibilities for future work. While $\alpha$-attractors provide a simple implementation of the inflationary sector, inflationary models that contain two or more distinct phases of inflation, like double inflation, side-tracked inflation and angular inflation, can help alleviate the parameter constraints of our current implementation and produce distinct GW features either at large or small scales. 
\textcolor{black}{A calculation of the effect of scalar fluctuations arising in the gauge-flation sector on the inflaton fluctuations can also provide interesting signatures, at the linear or non-linear level. This  also constitutes a necessary addition to the present analysis, as it can affect the resulting value of the tensor-to-scalar ratio $r$.}
Furthermore, inflationary models with non-Abelian gauge fields can have interesting consequences for baryogenesis and dark matter production \cite{Caldwell:2017chz, Adshead:2017znw, Maleknejad:2020yys, Maleknejad:2021nqi}, leading to correlated observables.

Furthermore, the original choice of the higher-order term for gauge-flation was based on the requirement for a vacuum energy-like equation of state $w\simeq -1$, required for driving inflation. Using an $SU(2)$ sector as a spectator sector opens up the possibility of introducing more non-linear terms, since the requirement of $w\simeq -1$ is lifted. It is interesting to explore the phenomenology of gauge-flation with other non-linear terms, dictated solely by the underlying symmetries, and their possible GW signatures. We leave this exploration for future work.

%%%%%%%%%%%%%%%%%%%%%%%%%%%%%%%%%%%%%%%%%%%%%%%%%%%%%%%%%%%%%%%%%%%%%%%%%%%%%%
%%%%%%%%%%%%%%%%%%%%%%%%%%%%%%%%%%%%%%%%%%%%%%%%%%%%%%%%%%%%%%%%%%%%%%%%%%%%%%%

\section*{Acknowledgements}
We thank P.~Adshead, T.~Fujita and A.~Maleknejad for useful discussions. The work of OI and EIS was supported by the
Netherlands Organization for Scientific Research (NWO).
The work of EIS was also supported  by a fellowship from ``la Caixa'' Foundation
(ID 100010434) and from the European Union's Horizon 2020 research and innovation programme under the Marie
Sk\l odowska-Curie grant agreement No 847648. The fellowship code is LCF/BQ/PI20/11760021.

%%%%%%%%%%%%%%%%%%%%%%%%%%%%%%%%%%%%%%%%%%%%%%%%%%%%%%%%%%%%%%%%%%%%%%%%%%%%%%
%%%%%%%%%%%%%%%%%%%%%%%%%%%%%%%%%%%%%%%%%%%%%%%%%%%%%%%%%%%%%%%%%%%%%%%%%%%%%%%

\appendix

\section{Blue-tilted GW spectrum}
\label{app:blueGW}

As shown in Eq.~\eqref{eq:QandnT}, the dynamics of $Q(t)$ controls the sign of the tensor tilt $n_T$. Here we present a realization where $Q(t)$ is an increasing function of time on the example of $\alpha$-attractor model, similarly as in Section~\ref{Sec:Viability}, for the same parameters of Eq.~\eqref{paramsV} for the potential, but with $\alpha=1$. The parameters we use for the gauge sector are
\begin{gather}
 g=1.7\times 10^{-2},\quad\kappa=10^{21}\, M_{\rm pl}^{-4}, \quad\dot{Q}_0/ M_{\rm pl}^2=10^{-10},\quad Q_0/ M_{\rm pl}=6\times10^{-4},
 8\times10^{-4}, 1.18\times10^{-3}.\label{params1}
\end{gather}

\begin{figure}
\centering
 \includegraphics[width=0.41\textwidth]{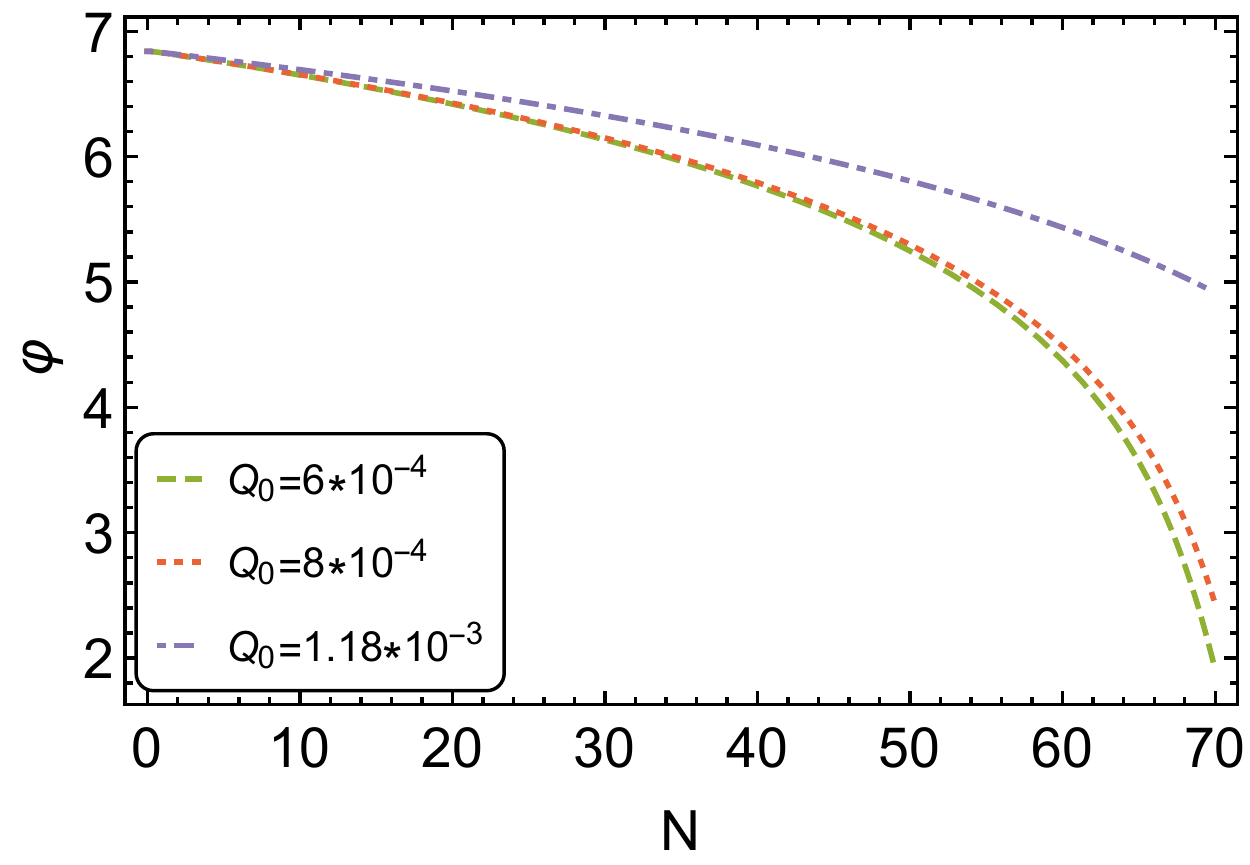}
\includegraphics[width=0.46\textwidth]{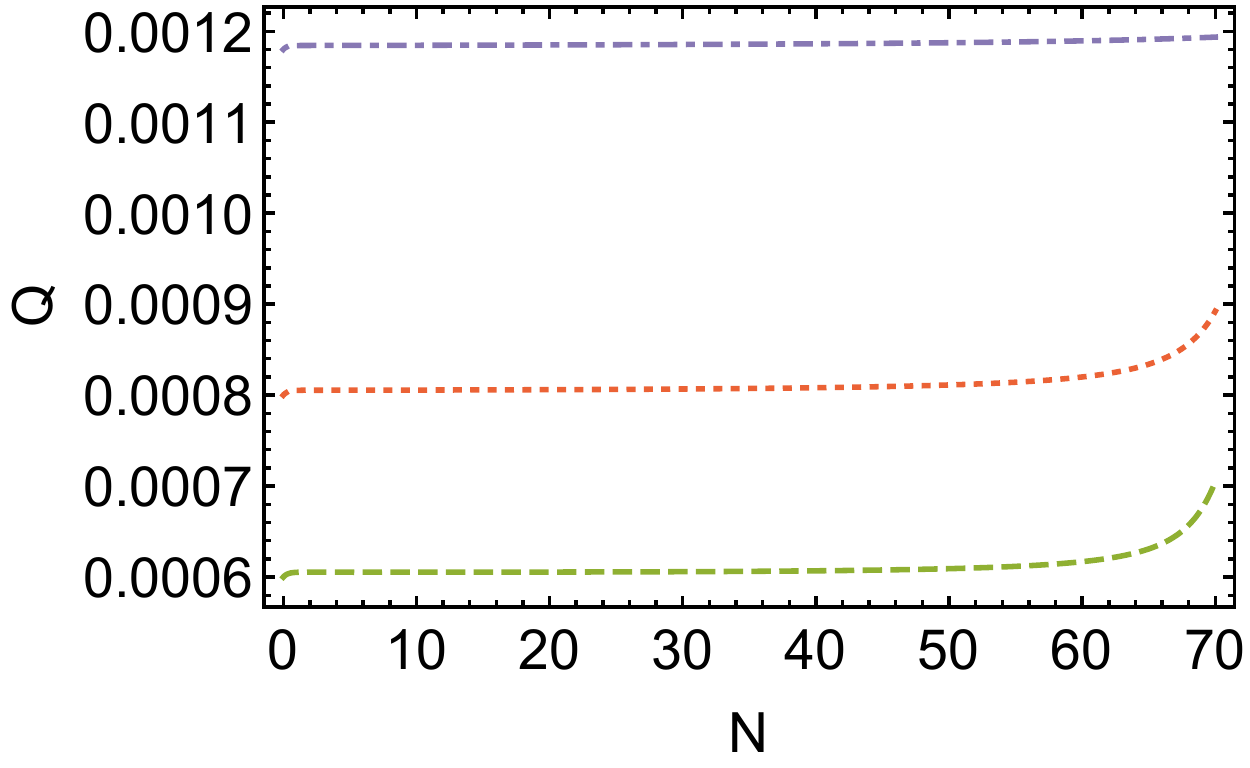}
\caption{
 {\it Left:} The dependence of  the inflaton field $\varphi$ on the $e$-folding number  $N$ for the $\alpha$-attractor T-model potential of Eq.~\eqref{Tpotential} for $Q_0/M_{\rm pl}=6\times10^{-4},
 8\times10^{-4}, 1.18\times10^{-3}$ (green-dashed, red-dotted and purple-dot-dashed lines respectively). 
 {\it Right:} The dependence of the  gauge field VEV $Q$  on the $e$-folding number  $N$ for the same potential and color coding.
 }
 \label{fig:PhiandQBlue}
\end{figure}

\begin{figure}
\centering
 \includegraphics[width=0.43\textwidth]{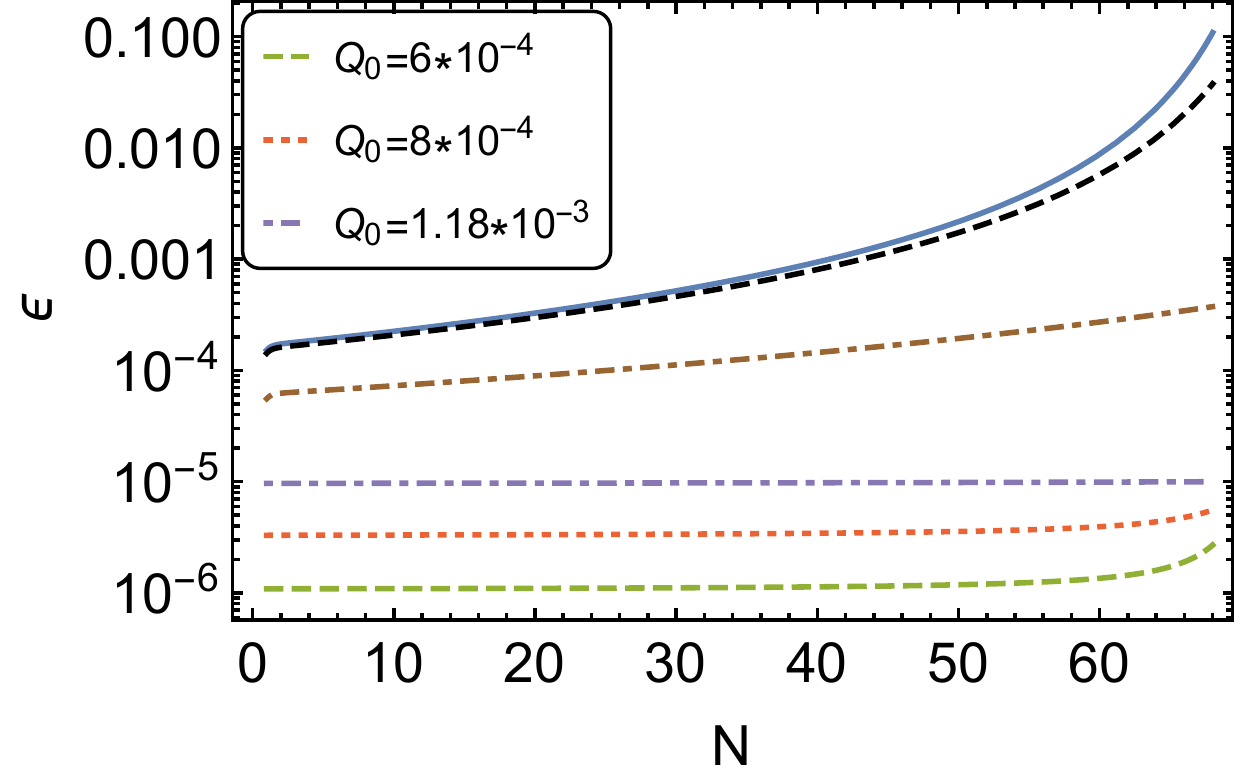}
\includegraphics[width=0.46\textwidth]{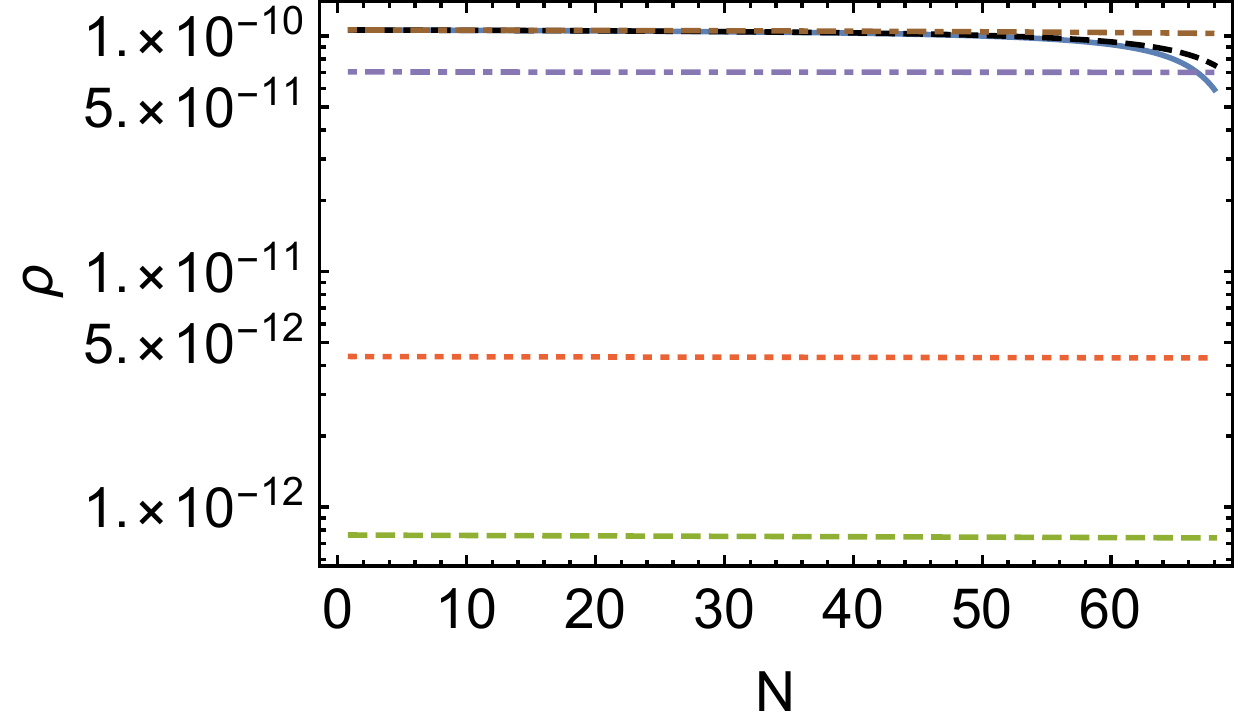}
\includegraphics[width=0.46\textwidth]{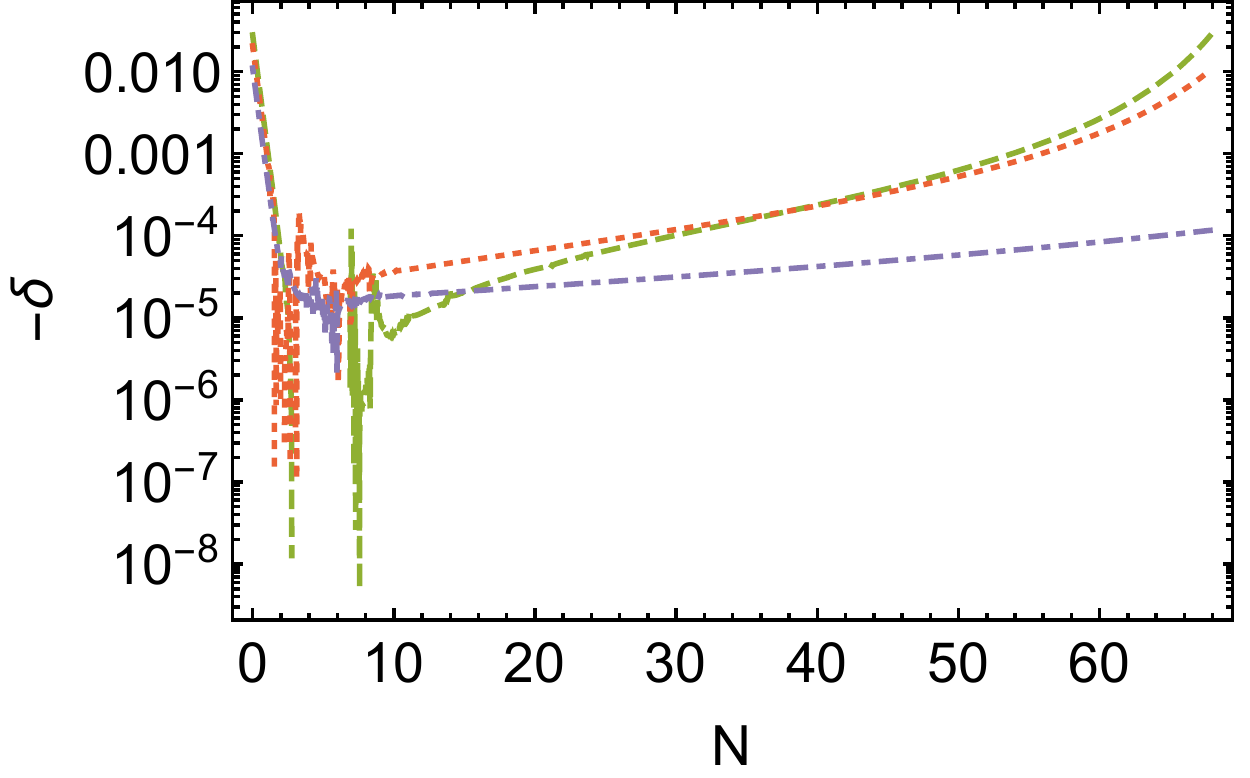}
\includegraphics[width=0.41\textwidth]{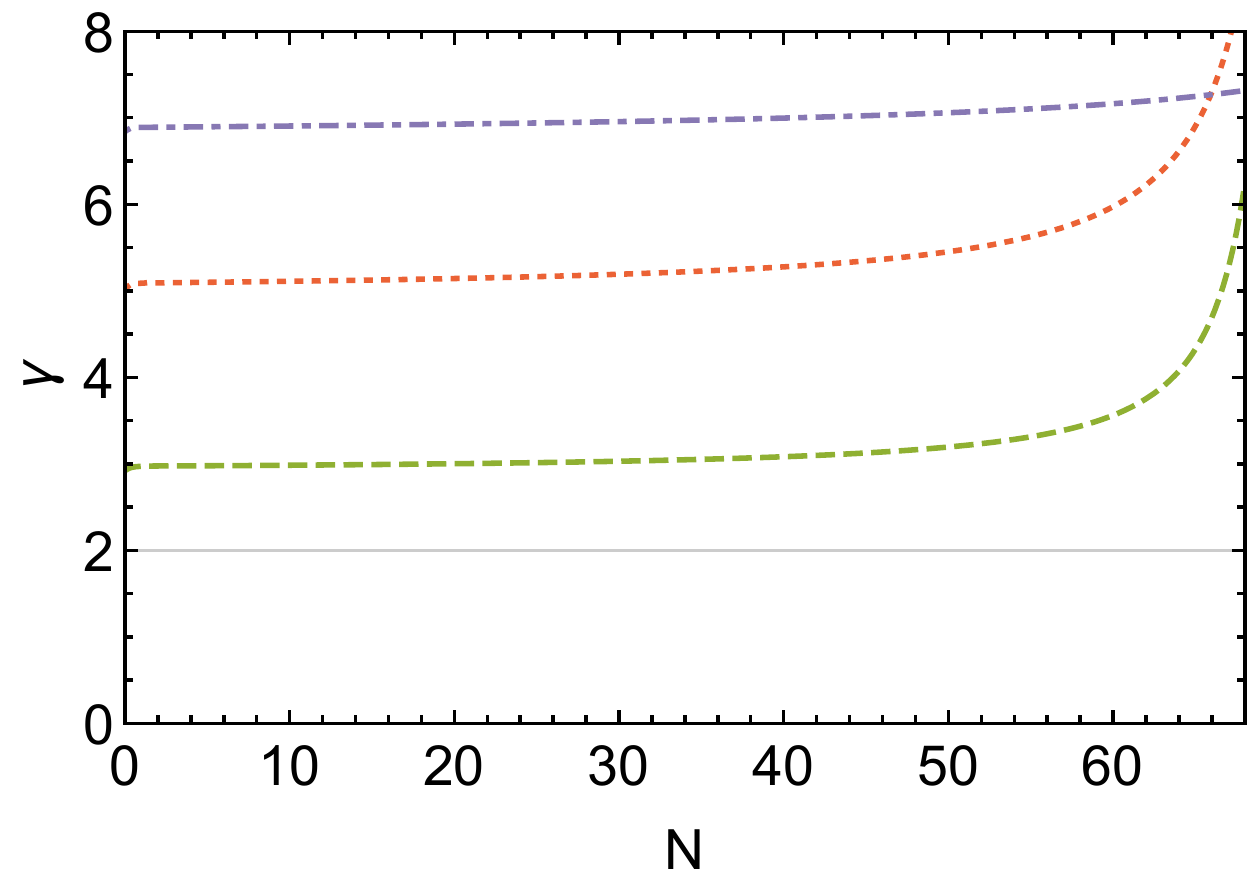}
\caption{
  {\it Top left:} Components $\epsilon_{Q_B}$ 
 as a function of the $e$-folding number $N$ 
for $Q_0/M_{\rm pl}=6\times10^{-4},
 8\times10^{-4}, 1.18\times10^{-3}$ (green-dashed, red-dotted and purple-dot-dashed lines respectively). 
   The blue-solid, black-dashed and brown-dot-dashed and  curved correspond to $\epsilon_{\varphi}$ for $Q_0/M_{\rm pl}=6\times10^{-4},
 8\times10^{-4}, 1.18\times10^{-3}$ respectively. 
   {\it Top right:} Components $\rho_{\kappa}$ and their dependence on $N$ for the same $Q_0$ and color-coding. The very top curves  correspond to $\rho_{\varphi}$ and are practically indistinguishable.
{\it Bottom row:} The evolution of the parameter $\delta$ (left) and $\gamma$ (right) for the same parameters and color-coding. The solid grey grid line on the right panel shows the bound $\gamma=2$, below which scalar fluctuations in the theory are unstable. 
 }
 \label{fig:EpsDeltaBlue}
\end{figure}

\begin{figure}
\centering
 \includegraphics[width=0.43\textwidth]{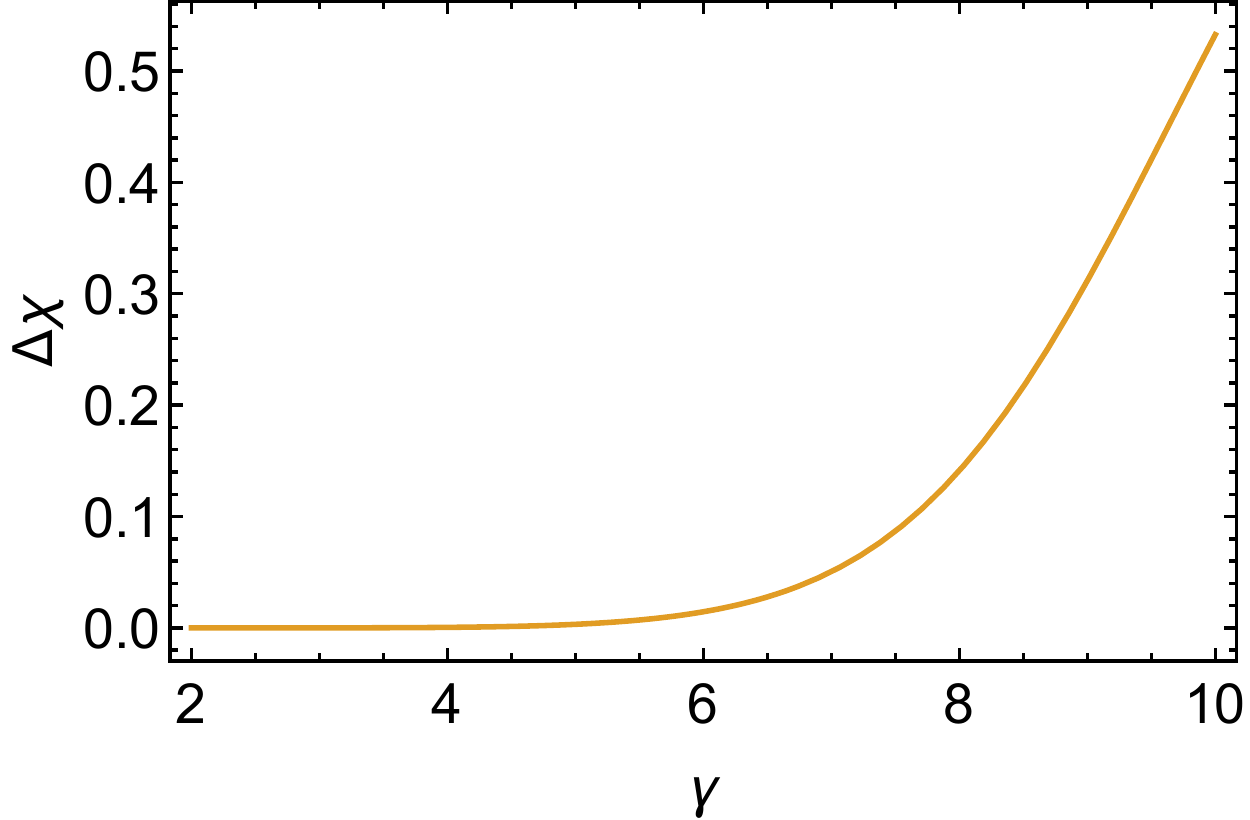}
\caption{The chirality parameter $\Delta \chi$ as a function of $\gamma$ for $\alpha=1$, $H/M_{\rm pl}=6*10^{-6}$ and $g=1.7*10^{-2}$. In the allowed region $\gamma\lesssim 7$ the chirality parameter is small.
 }
 \label{fig:ChiralityBlue}
\end{figure}

Our numerical simulations show that in order for $Q(t)$ to increase with time, one has to impose higher values of the parameter $\kappa$ in comparison with those used in Section~\ref{Sec:Viability}. Because of the conditions of Eqs.~\eqref{rho4} and \eqref{condit5}, this highly constrains the values of the allowed $Q_0$ and $g$, and hence via Eq.~\eqref{delta} limits the allowed range for the parameter $\gamma$ that controls the enhancement of chiral gravitational waves.

Fig.~\ref{fig:PhiandQBlue} shows the evolution of the inflaton field $\varphi$ and the vacuum expectation value of the gauge field $Q$ with the $e$-folding number  $N$, that behave similarly to those discussed in Section~\ref{Sec:Viability}, but with $Q$ being a slowly  increasing function of time. In such case $\delta$ becomes negative, as shown in Fig.~\ref{fig:EpsDeltaBlue}. However, one may see from the top right panel of the Fig.~\ref{fig:EpsDeltaBlue}, that further increase of $Q_0$ will violate the condition $\rho_{\varphi}\gg \rho_{Q_{\kappa}}$. Hence for the parameters of Eq.~\eqref{params1}, we compute $\gamma\simeq 7$ as the maximum value.
 From Fig.~\ref{fig:chirality} we can see that for this value the chirality parameter is only $\Delta \chi \simeq 0.05$, hence no significant production of sourced gauge fields has taken place. While this does not preclude the existence of a realization of this model, leading to significant $r$ and $n_T>0$, it demonstrates that it would require a significant level of parameter fine-tuning.

\end{document}